\pdfoutput=1

\documentclass[11pt]{article}

\usepackage[preprint]{acl}

\usepackage{times}
\usepackage{latexsym}

\usepackage[T1]{fontenc}

\usepackage[utf8]{inputenc}

\usepackage{microtype}

\usepackage{inconsolata}

\usepackage{graphicx}

\usepackage{booktabs}
\usepackage{subcaption}
\usepackage{multirow}
\usepackage{makecell}
\usepackage{enumitem}
\usepackage{pifont}
\usepackage{placeins}
\usepackage{stfloats}
\usepackage{amsmath}
\usepackage{amsfonts}
\usepackage{colortbl}
\usepackage{threeparttable}

\usepackage{CJKutf8}  

\usepackage[breakable]{tcolorbox}
\newtcolorbox{promptbox}{
    colback=gray!5,
    colframe=gray!75,
    boxrule=0.5pt,
    arc=4pt,
    left=6pt,
    right=6pt,
    top=6pt,
    bottom=6pt,
    breakable,
    fontupper=\small  
}

\definecolor{lightgreen}{rgb}{0.88, 1, 0.88}
\definecolor{lightred}{rgb}{1, 0.88, 0.88}
\definecolor{lightyellow}{rgb}{1, 1, 0.88}
\definecolor{lightblue}{rgb}{0.88, 0.95, 1}
\definecolor{lightorange}{rgb}{1, 0.93, 0.88}

\title{Advancing Zero-shot Text-to-Speech Intelligibility across\\Diverse Domains via Preference Alignment}

\author{
\vspace{-8mm}
\\
 \textbf{Xueyao Zhang}\textsuperscript{*,1},
 \textbf{Yuancheng Wang}\textsuperscript{*,1},
 \textbf{Chaoren Wang}\textsuperscript{1},\\
 \textbf{Ziniu Li}\textsuperscript{1}, 
 \textbf{Zhuo Chen}\textsuperscript{2},
 \textbf{Zhizheng Wu}\textsuperscript{1}
\\
 \textsuperscript{1}The Chinese University of Hong Kong, Shenzhen \\
 \textsuperscript{2}ByteDance Seed
}

\begin{document}
\maketitle

\begin{abstract}
Modern zero-shot text-to-speech (TTS) systems, despite using extensive pre-training, often struggle in challenging scenarios such as tongue twisters, repeated words, code-switching, and cross-lingual synthesis, leading to intelligibility issues. To address these limitations, this paper leverages preference alignment techniques, which enable targeted construction of out-of-pretraining-distribution data to enhance performance. We introduce a new dataset, named the \underline{Int}elligibility \underline{P}reference Speech Dataset (INTP), and extend the Direct Preference Optimization (DPO) framework to accommodate diverse TTS architectures. After INTP alignment, in addition to intelligibility, we observe overall improvements including naturalness, similarity, and audio quality for multiple TTS models across diverse domains. Based on that, we also verify the weak-to-strong generalization ability of INTP for more intelligible models such as CosyVoice 2 and Ints. Moreover, we showcase the potential for further improvements through iterative alignment based on Ints. Audio samples are available at \href{https://intalign.github.io/}{https://intalign.github.io/}.
\end{abstract}

\section{Introduction}

\renewcommand{\thefootnote}{*}
\footnotetext{Equal Contribution.}

\renewcommand{\thefootnote}{\arabic{footnote}} 
\addtocounter{footnote}{0}%


Despite leveraging large-scale pre-training~\cite{seedtts,maskgct,cosyvoice2}, modern zero-shot TTS systems still lack robustness during real-world applications~\cite{hallucination-survey,robustness-problem}. These systems struggle to meet even the most fundamental requirement of speech synthesis -- \textit{intelligibility}~\cite{tts-book-tanxu} in several scenarios, including: (1) the target text is hard to pronounce, such as tongue twisters or continuously repeated words~\cite{robustness-problem,seedtts}, which is referred to as \textit{articulatory} cases in this paper, (2) \textit{code-switching} cases, where the target text contains a mixture of multiple languages, and (3) \textit{cross-lingual} cases, where the languages of the target text and the reference speech differ. In these domains, existing zero-shot TTS models frequently exhibit ``hallucination'' issues, such as content insertion, omission, and mispronunciation~\cite{robustness-problem,valle}.

We attribute these intelligibility challenges primarily to the problem of \textit{out-of-distribution} (OOD). For example, in cross-lingual cases, there exists a huge mismatch between monolingual pre-training and cross-lingual inference. While including such scenarios in pre-training data would be a natural solution, collecting high-quality data for challenging cases like cross-lingual synthesis remains difficult. 

Motivated by the above, we propose to use \textit{preference alignment} (PA)~\cite{instructgpt,RLHF-anthoropic} to mitigate the OOD issues, and thus enhance zero-shot TTS intelligibility. The potential of this approach lies in two aspects. 
First, PA's \textit{customized} post-training on human expected distribution can effectively mitigate the OOD issue~\cite{zhang2024negative,li2024rl, iterative-dpo}. Second, unlike TTS pre-training that requires high-quality supervised data, PA needs only paired samples with relative preferences -- notably, even synthetic data can lead to large improvements~\cite{llama3,qwen2.5}, thus significantly simplifying data collection for challenging scenarios like cross-lingual cases. Centered on this direction, we investigate three research problems:
\vspace{-1mm}
\begin{itemize}[itemsep=0ex,leftmargin=2ex]
    \item \textbf{P1}: Dataset quality is crucial for model performance. To construct a high-quality intelligibility preference dataset, what prompts and base models should be selected, and how can we establish human-aligned preference pairs?
    \item \textbf{P2}: Unlike textual LLMs with predominantly autoregressive (AR) design, zero-shot TTS models employ diverse architectures, including AR-based~\cite{audiolm,seedtts,cosyvoice2}, Flow-Matching (FM) based~\cite{voicebox,e2tts,f5tts}, and Masked Generative Model (MGM) based~\cite{naturalspeech3,maskgct}. How can we design alignment algorithms for various architectures?
    \item \textbf{P3}: Can our preference dataset demonstrate weak-to-strong generalization~\cite{weak-to-strong-generalization}? In other words, can datasets created using less capable models effectively train more powerful models? This question is central to understanding the scalability and transferability of our dataset design.
\end{itemize}

In this paper, we address the aforementioned problems with the following key contributions:

$\rightarrow$ \textbf{P1}: We establish a synthetic \underline{Int}elligibility \underline{P}reference Speech Dataset (INTP), comprising about 250K preference pairs (over 2K hours) of diverse domains. Specifically, INTP covers multiple scenarios, utilizing various TTS models for data creation. Besides, we employ several strategies to construct preference pairs, aiming to mitigate the risk of reward hacking for simple patterns~\cite{reward-hacking,reward-hacking-wenglilian}. Particularly, we leverage human knowledge and DeepSeek-V3~\cite{deepseek-v3} to introduce perturbations into TTS systems, creating human-guided negative samples. In addition, when using Word Error Rate (WER) to determine intelligibility preferences, we not only consider self-comparison within a single model as in previous studies~\cite{tianjinchuan-pa,fpo,koel-tts}, but also introduce comparisons across different models to leverage their complementary capabilities.

$\rightarrow$ \textbf{P2}: We adopt the idea of Direct Preference Optimization (DPO)~\cite{dpo} to enhance various zero-shot TTS architectures. We employ the vanilla DPO algorithm for AR-based TTS models, while proposing extended versions of it for FM-based and MGM-based models. Our experiments on INTP shows that these algorithms effectively improve the intelligibility, naturalness, and overall quality of multiple state-of-the-art TTS systems, including ARS (AR-based)~\cite{maskgct}, F5-TTS (FM-based)~\cite{f5tts}, and MaskGCT (MGM-based)~\cite{maskgct}.


$\rightarrow$ \textbf{P3}: To investigate INTP's weak-to-strong generalization capability~\cite{weak-to-strong-generalization} on more powerful base models, we research its alignment effects on CosyVoice 2~\cite{cosyvoice2} and Ints (Appendix~\ref{sec:appendix-detail-ins}). Both models are initialized from textual LLMs (CosyVoice 2: from Qwen2.5, 0.5B~\cite{qwen2}. Ints: from Phi-3.5-mini-instruct, 3.8B~\cite{phi3}) and achieve superior intelligibility performance (Table~\ref{tab:expt-main}). Our experimental results verify that INTP, though constructed from weaker models, remains effective for these two strong models. Additionally, we showcase how to establish an \textit{iterative} preference alignment ``flywheel'' of data and model improvements~\cite{RLHF-anthoropic,llama3,iterative-dpo} based on Ints.

We open-source all resources used in this study at Amphion~\cite{amphion}, including: (1) the proposed INTP\footnote{\href{https://huggingface.co/datasets/amphion/INTP}{https://huggingface.co/datasets/amphion/INTP}} and DPO-based alignment codebase for various TTS models\footnote{\href{https://github.com/open-mmlab/Amphion}{https://github.com/open-mmlab/Amphion}}, (2) all the INTP-enhanced models\footnote{\href{https://huggingface.co/amphion/INTP}{https://huggingface.co/amphion/INTP}} based on Ints, CosyVoice 2, ARS, F5-TTS, and MaskGCT, and (3) our newly constructed zero-shot TTS evaluation sets across diverse domains\footnote{\href{https://huggingface.co/datasets/amphion/Amphion-TTS-Eval}{https://huggingface.co/datasets/amphion/Amphion-TTS-Eval}}.

\section{Related Work}

\paragraph{Zero-Shot Text to Speech}
Given a target text and a reference speech as input, zero-shot TTS systems aim to synthesize the target text while mimicking the reference style. Modern zero-shot TTS systems include AR approaches~\cite{valle, voicecraft, seedtts, fireredtts, cosyvoice, cosyvoice2, vevo} that model discrete speech tokens~\cite{soundstream, encodec}, and Non-AR approaches that either model continuous representations using diffusion~\cite{naturalspeech2} or flow matching~\cite{voicebox, e2tts, f5tts}, or model discrete tokens using masked generative models~\cite{soundstorm, naturalspeech3, maskgct, wang2025metis}. While these systems, trained on large-scale datasets~\cite{emilia, kahn2020libri,emilia-large}, show excellent intelligibility in regular cases~\cite{seedtts,librispeech,cosyvoice2}, they still struggle with intelligibility in real-world scenarios.

\vspace{-1mm}
\paragraph{Alignment for Speech Generation}
Alignment via post-training has demonstrated its effectiveness in the generation of text~\cite{instructgpt,RLHF-anthoropic}, vision~\cite{imagereward,tldr-perturb-vision}, speech~\cite{speechalign,seedtts,cosyvoice2}, music~\cite{musicrl}, and sound effects~\cite{tango2,baton}. In speech generation, existing works have employed preference alignment to enhance multiple aspects of speech, including intelligibility~\cite{seedtts,cosyvoice2,tianjinchuan-pa}, speaker similarity~\cite{seedtts,cosyvoice2,tianjinchuan-pa}, emotion controllability~\cite{seedtts,emo-dpo}, and overall quality~\cite{speechalign,uno,rio,dlpo,fpo,koel-tts}. 
For intelligibility, previous studies choose WER as the optimization objective, either directly employing it as a reward model~\cite{seedtts,cosyvoice2} or centering around it to construct preference pairs~\cite{tianjinchuan-pa,fpo,koel-tts}. 

However, the existing research exhibits two main limitations. First, in constructing intelligibility preference dataset, current works rely solely on a single model to generate data~\cite{tianjinchuan-pa,fpo,koel-tts}, neglecting comparisons across different models. Additionally, beyond the objective WER, the potential of leveraging human knowledge or feedback to construct preference pairs remains unexplored. Second, most existing work has focused primarily on optimizing AR-based~\cite{speechalign,seedtts,cosyvoice2,tianjinchuan-pa} or diffusion-based~\cite{dlpo} TTS models, leaving open the question of how to design effective alignment algorithms for other architectural paradigms, such as FM-based and MGM-based TTS models.

\section{INTP: Intelligibility Preference Speech Dataset}

\begin{table*}[t]
\begin{center}
\begin{threeparttable}
    \begin{minipage}{0.68\textwidth}
    \subfloat[Distribution of preference pairs, where pronunciation-perturbed and punctuation-perturbed texts are introduced to create the human-guided negative samples.\label{tab:intp-stats}]{
    \resizebox{\textwidth}{!}{
    \begin{tabular}{r|ccccc|c}
        \toprule
         & \textbf{\makecell[c]{Regular}} & \textbf{\makecell[c]{Repeated}} & \textbf{\makecell[c]{Code-Switching}} & \textbf{\makecell[c]{Pronunciation-\\perturbed}} & \textbf{\makecell[c]{Punctuation-\\perturbed}} & \textbf{\#Total} \\
        \midrule
        \textbf{ARS}~\cite{maskgct} & 8,219 & 8,852 & 8,300 & 7,325 & 8,036 & 40,732 \\
        \textbf{F5-TTS}~\cite{f5tts} & 8,425 & 8,555 & 7,976 & 7,909 & 6,667 & 39,532 \\
        \textbf{MaskGCT}~\cite{maskgct} & 9,055 & 10,263 & 8,289 & 7,604 & 7,686 & 42,897 \\
        \midrule
        \textbf{Intra Pairs} & 25,699 & 27,670 & 24,565 & 22,838 & 22,389 & 123,161 \\
        \textbf{Inter Pairs} & 27,008 & 27,676 & 24,651 & 25,045 & 23,970 & 128,350 \\
        \midrule
        \textbf{\#Total} & 52,707 & 55,346 & 49,216 & 47,883 & 46,359 & 251,511 \\
        \bottomrule
    \end{tabular}
    }
    }
    \end{minipage}%
    \hfill
    \begin{minipage}{0.3\textwidth}
    \subfloat[Examples of different types for a text, ``\textit{A panda eats shoots and leaves}''.\label{tab:intp-examples}]{
    \resizebox{\textwidth}{!}{
    \begin{tabular}{r|l}
    \toprule
     \textbf{Text Type} & \textbf{Example} \\
     \midrule
     \textbf{Regular} & \textit{A panda eats shoots and leaves.} \\ \cmidrule(lr){1-2}
     \textbf{Repeated} & \textit{\makecell[l]{A panda panda eats shoots and\\leaves and leaves and leaves.}} \\ \cmidrule(lr){1-2}
     \textbf{Code-Switching} & 
        \begin{CJK*}{UTF8}{gbsn}
            \textit{熊猫吃~shoots~和~leaves。}
        \end{CJK*} \\ \cmidrule(lr){1-2}
    \textbf{\makecell[r]{Pronunciation-\\perturbed}} & \textit{A pan duh eights shots n leafs.} \\ \cmidrule(lr){1-2}
    \textbf{\makecell[r]{Punctuation-\\perturbed}} & \textit{A panda eats, shoots, and leaves.} \\
    \bottomrule
    \end{tabular}
    }
    }
    \end{minipage}
    \vspace{-2mm}
\end{threeparttable}
\caption{Intelligibility Preference dataset (INTP). There are about 250K pairs (over 2K hours) in INTP, covering various texts and speechs, multiple models, and diverse preference pairs.}
\label{tab:intp}
\vspace{-3mm}
\end{center}
\end{table*}


To enhance the TTS intelligibility, this study opts for constructing a preference dataset to align~\cite{tianjinchuan-pa,fpo,koel-tts} rather than directly optimizing single metrics or rules such as WER~\cite{seedtts,cosyvoice2}. This choice is motivated by two key considerations. First, through the construction of a preference dataset, we can inject human knowledge and feedback beyond WER, such as creating human-guided negative samples in the framework of preference alignment (Section~\ref{sec:human-guided-negative}). Second, in addition to the existing approach of constructing preference pairs from multiple samples of a single model~\cite{tianjinchuan-pa,fpo,koel-tts}, we can leverage comparisons across different models to create preference pairs, thereby utilizing the complementary capabilities of various models (Figure~\ref{fig:inter-pair}). These different strategies help increase diversity in the dataset, mitigating the risk of ``reward hacking'' that often results from the simple patterns inherent in single metrics or rules~\cite{RLHF-anthoropic,reward-hacking,reward-hacking-wenglilian}.


Formally, we aim to construct an intelligibility preference dataset $\mathcal{D} = \{(x, y^w, y^l)\}$, where each triplet comprises a prompt $x$ (consisting of target text $x^{text}$ and reference speech $x^{speech}$ for zero-shot TTS models), along with a pair of synthesized speech samples $(y^w, y^l)$. Here, $y^w$ and $y^l$ represent the preferred (positive) and dispreferred (negative) outputs conditioned on $x$, respectively. Statistics of the proposed INTP are presented in Table~\ref{tab:intp}. 

\subsection{Prompt Construction}\label{sec:prompt-construction}

To establish a high-quality preference dataset, we aim to make the distribution of prompt $x$ cover a wide range of domains. For the target text $x^{text}$, from the linguistic perspective, we design three distinct categories: (1) \textbf{Regular text}, which represents the general cases for TTS systems, aimed at enhancing model intelligibility in common scenarios; (2) \textbf{Repeated text}, which contains repeated or redundant words and phrases, specifically designed to improve TTS performance in articulatory cases; and (3) \textbf{Code-switching text}, which incorporates a mixture of different languages, intended to enhance TTS capabilities in multilingual scenarios. From the semantic perspective, we collect text content across diverse topics and domains to enrich the distribution of $x^{text}$. For the reference speech $x^{speech}$, we aim to cover a wide range of speakers, speaking styles, and acoustic environments. Regarding the pairing of $x^{text}$ and $x^{speech}$, we further consider their language alignment by constructing both \textbf{monolingual} and \textbf{cross-lingual} combinations (more statistics in Appendix~\ref{sec:appendix-prompt-construction}).

We construct these prompt data based on the Emilia-Large~\cite{emilia,emilia-large}, which contains real-world speech data and textual transcriptions across diverse topics, scenarios, and speaker styles. We perform stratified sampling on Emilia-Large's speech and text data to obtain multilingual prompts. We employ DeepSeek-V3~\cite{deepseek-v3} to preprocess the sampled text, including typo correction, and use it as regular text. Based on these regular texts, we further utilize DeepSeek-V3 to transform them into different text types (as shown in Table~\ref{tab:intp-examples}). Construction details are provided in Appendix~\ref{sec:appendix-prompt-construction}.

\subsection{Model Selection}

\begin{figure}[t]
    \centering
    \begin{subfigure}[b]{0.49\columnwidth}
        \centering
        \includegraphics[width=\textwidth]{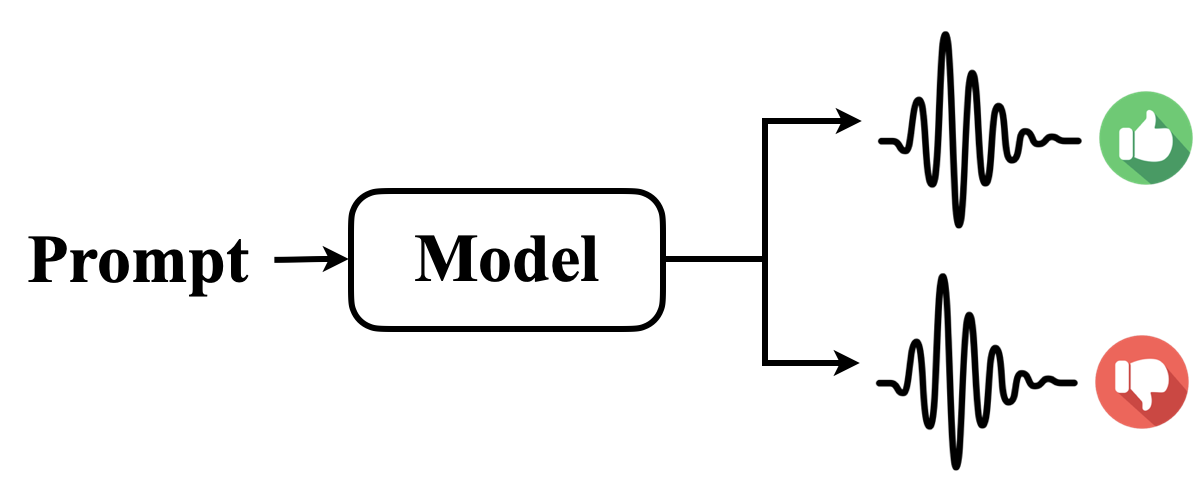}
        \caption{Intra Pair}
        \label{fig:intra-pair}
    \end{subfigure}
    \hfill
    \begin{subfigure}[b]{0.49\columnwidth}
        \centering
        \includegraphics[width=\textwidth]{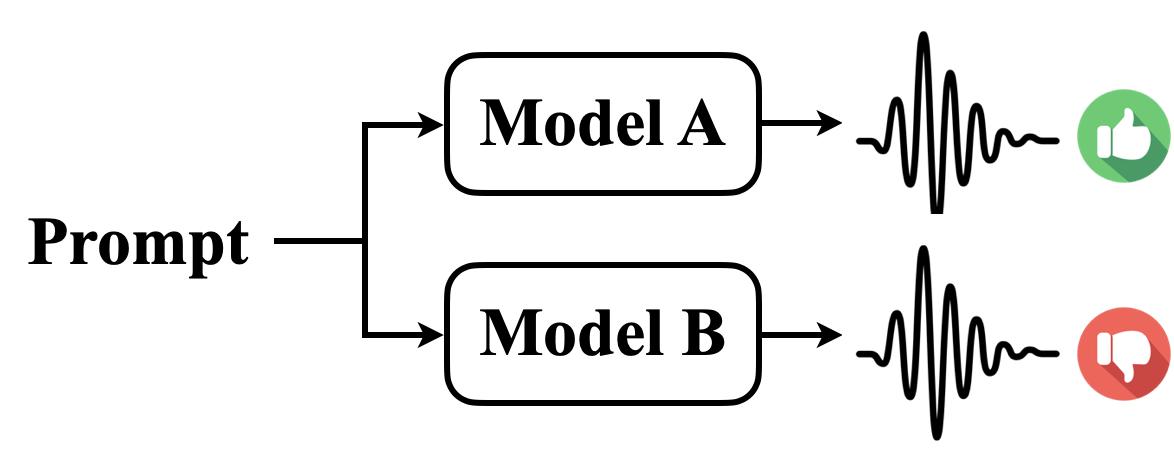}
        \caption{Inter Pair}
        \label{fig:inter-pair}
    \end{subfigure}
    \hfill
    \begin{subfigure}[b]{0.49\columnwidth}
        \centering
        \includegraphics[width=\textwidth]{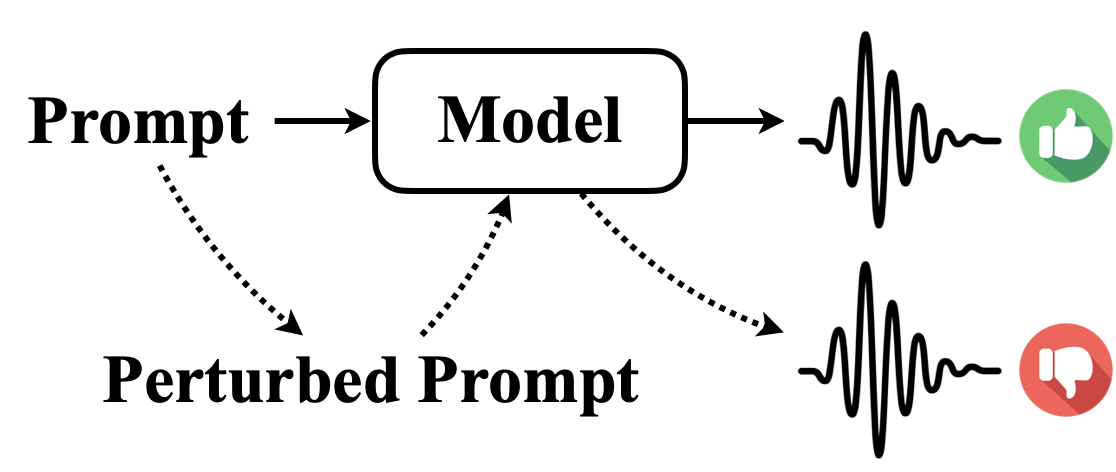}
        \caption{Perturbed Pair}
        \label{fig:perturbed-pair}
        \vspace{-3mm}
    \end{subfigure}
    \caption{Three kinds of preference pairs in INTP.}
    \label{fig:preference-pairs}
    \vspace{-5mm}
\end{figure}

We utilize multiple zero-shot TTS models with diverse architectures for data synthesis to enhance INTP's diversity and generalization. Specifically, we select the following three models:
(1) \textbf{ARS} (AR-based): Introduced as an autoregressive baseline by~\citet{maskgct}. and referred to as ``AR + SoundStorm'' in the original paper~\cite{maskgct}. It adopts a cascaded architecture, including the autoregressive \textit{text-to-codec} and the non-autoregressive \textit{codec-to-waveform}~\cite{soundstorm}.
(2) \textbf{F5-TTS} (FM-based): It follows E2 TTS~\cite{e2tts} and uses a flow-matching transformer~\cite{voicebox,flow-matching} to convert the text to acoustic features directly~\cite{f5tts}.
(3) \textbf{MaskGCT} (MGM-based): Similar to ARS, MaskGCT employs a two-stage architecture. The key distinction lies in its use of an MGM in the text-to-codec stage~\cite{maskgct}.

All the three are pre-trained on Emilia~\cite{emilia} (about 100K hours of multilingual data) and represent state-of-the-art zero-shot TTS systems across different architectures. We utilize their officially released pre-trained models (see Appendix~\ref{sec:appendix-model-selection} for details) to generate data for INTP.

\subsection{Preference Pairs Construction}\label{sec:preference-pairs-construction}

In constructing intelligibility preference pairs, we design three categories of pairs (Figure~\ref{fig:preference-pairs}):

\paragraph{Intra Pair} 
These pairs are generated through model self-comparison (Figure~\ref{fig:intra-pair}), following an approach similar to previous studies~\cite{tianjinchuan-pa,fpo,koel-tts}. For a given prompt $x$, we conduct multiple samplings using the same model. Subsequently, we calculate the WER for each generation and designate the samples with the lowest and highest WER as $y^w$ and $y^l$, respectively. To enlarge the gap between $y^w$ and $y^l$, we employ diverse sampling hyperparameters across multiple generations from the same model. Additionally, we use a specific WER threshold to filter out pairs with insufficient performance gaps (more details in Appendix~\ref{sec:appendix-intp-intra-pair}).

\paragraph{Inter Pair} 
These pairs are constructed by comparing outputs across different models (Figure~\ref{fig:inter-pair}). The efficacy of this approach lies in leveraging the complementary strengths of various models. For example, by comparing intra-pairs from different models for the same prompt, we can identify the ``best of the best'' samples, thereby enhancing the overall quality of positive samples in our dataset. Similar to intra pair, we also employ WER to identify intelligibility preferences for inter pairs (see Appendix~\ref{sec:appendix-intp-inter-pair} for details).

Notably, the proposed inter-pair construction pipeline enables comparative evaluation of intelligibility performance across different models. Using this pipeline, we compared four state-of-the-art models in the field: ARS~\cite{maskgct}, F5-TTS~\cite{f5tts}, MaskGCT~\cite{maskgct}, and CosyVoice 2~\cite{cosyvoice2}. We constructed 10K inter-pairs and analyzed the win rates of these models, as shown in Table~\ref{tab:intp-inter-arena}. Interestingly, even ARS, the model with the lowest win rate, achieves a 4.1\% success rate against the strongest model, CosyVoice 2. This finding validates our assumption regarding the complementary capabilities among various models.

\paragraph{Perturbed Pair}\label{sec:human-guided-negative}
In addition to the aforementioned two types of pairs which are established based on WER, we leverage human knowledge and DeepSeek-V3~\cite{deepseek-v3} to create human-guided negative samples, termed perturbed pairs (Figure~\ref{fig:perturbed-pair}). The main idea involves deliberately perturbing the input prompt, thereby inducing the model to generate low-quality samples~\cite{tango2,tldr-perturb-vision}. 

We design two types of perturbation for the target text in the prompt (as shown in Table~\ref{tab:intp-examples}): (1) \textbf{Pronunciation perturbation}: we replace certain characters of the text with easily mispronounceable alternatives. For example, given the text ``\textit{A panda eats shoots and leaves}'', we can create the perturbed text ``\textit{A pan duh eights shots n leafs}''. (2) \textbf{Punctuation perturbation}: we modify the punctuation, such as commas, to alter pause patterns and prosody in the text. For example, by adding commas to the text ``\textit{A panda eats shoots and leaves}'', we obtain ``\textit{A panda eats, shoots, and leaves}'', where the words ``\textit{shoots}'' and ``\textit{leaves}'' transform from nouns in the original text to verbs, creating a significant semantic shift. The detailed construction process is provided in Appendix~\ref{sec:appendix-intp-perturbed-pair}.

\subsection{Human Perception Verification}\label{sec:intp-human-verify}

\begin{table}[t]
\begin{center}
\resizebox{\columnwidth}{!}{%
\begin{threeparttable}
    \begin{tabular}{c|cccc|c}
        \toprule
         & \textbf{ARS} & \textbf{F5-TTS} & \textbf{MaskGCT} & \textbf{CosyVoice 2} & \textbf{Win Rate} ($\uparrow$) \\
        \midrule
        \textbf{ARS} & / & 6.7\% & 7.4\% & 4.1\% & 18.3\% \\
        \textbf{F5-TTS} & 10.4\% & / & 8.8\% & 5.9\% & 25.1\% \\
        \textbf{MaskGCT} & 10.4\% & 8.0\% & / & 5.9\% & 24.3\% \\
        \textbf{CosyVoice 2} & 11.9\% & 10.2\% & 10.3\% & / & 32.3\% \\
        \bottomrule
    \end{tabular}
    \begin{tablenotes}
        {
        \item[*] The percentage in each cell represents the proportion of cases where the model on the horizontal axis outperforms the model on the vertical axis. 
        \item[*] The \textbf{Win Rate} is calculated as the sum of values from columns 2 through 5.
        }
    \end{tablenotes}
\end{threeparttable}
}
\caption{TTS Intelligibility Arena: We employ the inter-pair construction from INTP to compare intelligibility among four state-of-the-art zero-shot TTS models.}
\label{tab:intp-inter-arena}
\vspace{-4mm}
\end{center}
\end{table}

\begin{table}[t]
\begin{center}
\resizebox{\columnwidth}{!}{%
\begin{threeparttable}
    \begin{tabular}{c|cccc}
        \toprule
         & \makecell[c]{\textbf{ARS}} & \makecell[c]{\textbf{F5-TTS}} & \makecell[c]{\textbf{MaskGCT}} & \makecell[c]{\textbf{CosyVoice 2}} \\
         \midrule
         \textbf{Positive Samples} & 73.0\% & 88.1\% & 90.9\% & 100.0\% \\
         \textbf{Negative Samples} & 45.7\% & 15.8\% & 47.1\% & 75.0\% \\
         \midrule
         \textbf{All} & 59.7\% & 53.7\% & 64.3\% & 90.4\% \\
         \bottomrule
    \end{tabular}
\end{threeparttable}
}
\caption{Human-annotated reading accuracy ($\uparrow$) for four state-of-the-art zero-shot TTS models on regular texts. We use the intra-pair pipeline of INTP to generate the positive and negative samples.}
\label{tab:human-read-acc-verify}
\vspace{-5mm}
\end{center}
\end{table}

After constructing INTP, we further conducted subjective evaluation to verify its alignment with human perception. For intelligibility alignment, we design a \textbf{reading accuracy} listening task (see Appendix~\ref{sec:appendix-sub-eval} for details): given a text and a speech, subjects perform binary classification to determine whether the speech accurately reads the text without any content insertion, omission, or mispronunciation. Using four state-of-the-art zero-shot TTS models, we generate 300 intra-pairs on INTP regular texts. The results in Table~\ref{tab:human-read-acc-verify} demonstrate that INTP's preference identification for intra pairs aligns well with human judgments of intelligibility. Furthermore, comparing Tables~\ref{tab:intp-inter-arena} and \ref{tab:human-read-acc-verify} reveals that INTP's inter-pair comparisons of intelligibility across different models also effectively align with human values.

In addition to intelligibility, we also investigated how well INTP aligns with human preferences for \textit{naturalness}, which is one of the most general-purpose metrics for TTS~\cite{tts-book-tanxu}. The experimental results demonstrate that the naturalness gap between positive and negative samples of INTP is substantial and perceptible to human listeners. We discuss this finding in details in Appendix~\ref{sec:appendix-intp-human-verficiation}.

\section{Preference Alignment for Diverse Zero-Shot TTS models}\label{method:dpo}

In this section, we present methods for achieving preference alignment across a range of TTS models, including autoregressive based, flow-matching based, and masked generative model based architectures. Building on the framework of Direct Preference Optimization (DPO)~\cite{dpo}, initially developed for AR-based models, we adapt and extend its principles to FM-based and MGM-based models. We note that DPO is computationally efficient in practice, and its iterative variant aligns seamlessly with the online reinforcement learning (RL) framework \citep{li2023rema4}.



\subsection{DPO for AR Models}
\label{method:dpo_ar}

The main idea of reinforcement learning (RL) for preference alignment is to introduce a reward model  $r(x,y)$ to guide the model for improvement (see e.g., \citep{li2023rema4}). Here $y$ represents the output (i.e., the generated speech in zero-shot TTS), and $x$ means the input prompt (i.e., the reference speech and the target text in zero-shot TTS).
A widely adopted reward model design is based on Bradley-Terry (BT) model, which defines the probability of preferred sample $y^w$ over dispreferred sample $y^l$ given $x$ as $p_\text{BT}(y^w \succ y^l \mid x) = \sigma (r(x, y^w) - r(x, y^l))$.
We can train the reward model \( r_\phi(x, y) \) by minimizing the negative log-likelihood of observed comparisons from the preference dataset $\mathcal{D}$:
\begin{equation}
    {\small
    \begin{aligned}
 \mathcal{L}_\text{R} = -\mathbb{E}_{(x, y_w, y_l) \sim \mathcal{D}} \left[ \log \sigma\left(r_\phi(x, y_w) - r_\phi(x, y_l)\right) \right].
    \end{aligned}
    }
\label{eq:reward_model}
\end{equation}
With the given reward model, the RL optimization objective is to guide the model to maximize the expected reward while minimizing the KL-divergence from a reference distribution:
\begin{equation}
    {\small
    \begin{aligned}
  \max_{p_\theta} \mathbb{E}_{x, y \sim p_\theta(y | x)} [r(x, y)] - \beta D_{\text{KL}} [p_\theta(y | x) \parallel p_{\text{ref}}(y | x)],  
  \end{aligned}
  }
\label{eq:rl_obj}
\end{equation}
where the hyperparameter $\beta$ controls the strength of the regularization.
As highlighted in~\citet{dpo}, the optimization problem in Equation~\ref{eq:rl_obj} admits a closed form solution. This implies a direct relationship between the reward function and the policy. Substituting the reward expression into Equation~\ref{eq:reward_model} leads the DPO loss:
\begin{equation}
    {\small
    \begin{aligned}
\mathcal{L}_\text{DPO} = -\mathbb{E}_{\mathcal{D}} \left[ \log \sigma\left( \beta \left( \log \tfrac{p_\theta(y_w | x)}{p_{\text{ref}}(y_w | x)} - \log \tfrac{p_\theta(y_l | x)}{p_{\text{ref}}(y_l | x)} \right) \right) \right].
    \end{aligned}
    }
\label{eq:dpo_ar}
\end{equation}


DPO enables direct preference alignment for AR-based TTS models, eliminating the need for explicit reward modeling or RL optimization. In the following subsections, we will introduce its extensions for FM-based and MGM-based TTS models.

\subsection{DPO for Flow-Matching Models}
\label{method:dpo_fm}

The vanilla DPO algorithm is tailored for AR models, while~\citet{wallace2024diffusion} extends it to diffusion models. In this subsection, we introduce the DPO algorithm for flow-matching models, specifically demonstrating its application to optimal transport flow-matching (OT-FM), a common approach in FM-based TTS models~\cite{voicebox,e2tts,f5tts}.
Given the continuous representation $y$ of a speech sample and its corresponding condition $x$, OT-FM constructs a linear interpolation path between Gaussian noise $y_0 \sim \mathcal{N}(0, I)$ and the target data $y_1 = y$. Specifically, the interpolation follows $y_t = (1-t)y_0 + t\,y_1$, where $t \in [0,1]$,
which naturally induces a velocity field \( v_\theta(y_t, t, x) \) that captures the constant directional derivative
$\frac{dy_t}{dt} = y_1 - y_0$.
OT-FM aims to learn the velocity field to match the true derivative. The corresponding loss function is defined as
\begin{equation}
{\small
\begin{aligned}
\mathcal{L}_{\text{OT-FM}} = \mathbb{E}_{y_0, y_1, x, t}
\|\,v_\theta(y_t,t,x) - \left(y_1 - y_0\right)\|_2^2,
\end{aligned}
}
\label{eq:obj_fm}
\end{equation}
where $t$ is the time step that is sampled from the uniform distribution $\mathcal{U}(0, 1)$.

Inspired by~\citet{wallace2024diffusion}, we rewrite the RL objective for flow-matching models. Let $p_\theta(y_1|y_t,t,x)$ denote our policy that predicts the target sample $y_1$ given the noised observation $y_t$ at time $t$ and condition $x$. We initialize from a reference flow-matching policy $p_{\text{ref}}$. The RL objective can be written as:
\begin{equation}
{\small
\begin{aligned}
\max_{p_\theta} & \mathbb{E}_{y_1 \sim p_\theta(y_1|x), t, x}[r(y_1,x)] \\
& - \beta \mathbb{D}_{\text{KL}}[p_\theta(y_1|y_t,t,x) \| p_{\text{ref}}(y_1|y_t,t,x)].
\end{aligned}
}
\label{eq:rl_obj_fm}
\end{equation}
Following a similar derivation process as in DPO (we provide more details in Appendix~\ref{sec:appendix-derivation-fm}), we can obtain the loss function for flow-matching DPO:
\begin{equation}
{\scriptsize
\begin{aligned}
& \mathcal{L}_{\text{DPO-FM}} = -
\mathbb{E}_{(y_1^w, y_1^l, x) \sim \mathcal{D}, t} \\
& \log \sigma \left( \beta \left( \log \frac{p_\theta(y_1^w|y_t^w,t,x)}{p_{\text{ref}}(y_1^w|y_t^w,t,x)}
- \log \frac{p_\theta(y_1^l|y_t^l,t,x)}{p_{\text{ref}}(y_1^l|y_t^l,t,x)} \right) \right),
\end{aligned}
}
\label{eq:dpo-fm}
\end{equation}
where $y_1^w$ and $y_1^l$ represent the preferred and dispreferred samples from the preference dataset, respectively, while $y_t^w$ and $y_t^l$ are the interpolations at time $t$ between $y_1^w$ and $y_1^l$ and the randomly sampled $y_0^w$ and $y_0^l$. The loss can be transformed into the velocity space:
\vspace{-2mm}
\begin{equation}
{\tiny
\begin{split}
& \mathcal{L}_{\text{DPO-FM}} = -\mathbb{E}_{(y_1^w, y_1^l, x) \sim \mathcal{D}, t} \log \sigma \Big( -\beta \\
& \Big( \left\| v_\theta(y^w_t, t, x) - (y_1^w - y_0^w) \right\|_2^2
- \left\| v_\text{ref}(y^w_t, t, x) - (y_1^w - y_0^w) \right\|_2^2 \Big) \\
& - \Big( \left\| v_\theta(y^l_t, t, x) - (y_1^l - y_0^l) \right\|_2^2
- \left\| v_\text{ref}(y^l_t, t, x) - (y_1^l -y_0^l) \right\|_2^2 \Big) \Big).
\end{split}
}
\label{eq:dpo-fm-velocity}
\end{equation}

This proposed algorithm can be applied to a wide range of FM-based and diffusion-based TTS models~\cite{voicebox, e2tts, naturalspeech2}. In this study, we use it to optimize F5-TTS~\cite{f5tts} as a representative.

\begin{table*}[t]
\begin{center}
\begin{threeparttable}
    \resizebox{\textwidth}{!}{
    \begin{tabular}{l||rrc|rrc|rrc|rrc||rrc}
    \toprule
    \multirow{2}{*}{\textbf{Model}} & 
    \multicolumn{3}{c|}{\textbf{Regular cases}} &
    \multicolumn{3}{c|}{\textbf{Articulatory cases}} &
    \multicolumn{3}{c|}{\textbf{Code-switching cases}} &
    \multicolumn{3}{c||}{\textbf{Cross-lingual cases}} &
    \multicolumn{3}{c}{\textbf{Avg}} \\
    \cmidrule(lr){2-13} \cmidrule(lr){14-16}
     & \textbf{WER} & \textbf{SIM} & \textbf{N-CMOS} & \textbf{WER} & \textbf{SIM} & \textbf{N-CMOS} & \textbf{WER} & \textbf{SIM} & \textbf{N-CMOS} & \textbf{WER} & \textbf{SIM} & \textbf{N-CMOS} & \textbf{WER} & \textbf{SIM} & \textbf{N-CMOS} \\
    \midrule
    \multicolumn{1}{l||}{\textbf{ARS}}
        & 3.96 & 0.717 & -
        & 20.03 & 0.693 & - 
        & 54.15 & 0.693 & - 
        & 19.76 & 0.630 & - 
        & 24.47 & 0.683 & - \\
    \multicolumn{1}{r||}{\textbf{w/ INTP}}
        & 2.32 & 0.727 & 0.47 $_{\scriptscriptstyle \pm \text{0.22}}$
        & 12.83 & 0.713 & 0.64 $_{\scriptscriptstyle \pm \text{0.31}}$
        & 36.91 & 0.698 & 0.63 $_{\scriptscriptstyle \pm \text{0.34}}$
        & 9.57  & 0.632 & 0.82 $_{\scriptscriptstyle \pm \text{0.28}}$
        & 15.41 & 0.692 & 0.64 $_{\scriptscriptstyle \pm \text{0.12}}$ \\
    \midrule
    \multicolumn{1}{l||}{\textbf{F5-TTS}}
        & 3.44 & 0.670 & -
        & 16.84 & 0.635 & -
        & 33.99 & 0.609 & -
        & 16.86 & 0.546 & - 
        & 17.78 & 0.615 & - \\
    \multicolumn{1}{r||}{\textbf{w/ INTP}}
        & 2.38 & 0.652 & 0.38 $_{\scriptscriptstyle \pm \text{0.26}}$
        & 12.97 & 0.628 & 0.30 $_{\scriptscriptstyle \pm \text{0.23}}$
        & 15.98 & 0.576 & 0.67 $_{\scriptscriptstyle \pm \text{0.36}}$
        & 7.13 & 0.509 & 0.47 $_{\scriptscriptstyle \pm \text{0.30}}$
        & 9.62 & 0.591 & 0.44 $_{\scriptscriptstyle \pm \text{0.12}}$ \\
    \midrule
    \multicolumn{1}{l||}{\textbf{MaskGCT}}
        & 2.34 & 0.738 & - 
        & 12.43 & 0.714 & - 
        & 29.06 & 0.696 & - 
        & 12.34 & 0.629 & - 
        & 14.04 & 0.694 & - \\
    \multicolumn{1}{r||}{\textbf{w/ INTP}}
        & 2.23 & 0.737 & 0.23 $_{\scriptscriptstyle \pm \text{0.20}}$
        & 9.13 & 0.722 & 0.57 $_{\scriptscriptstyle \pm \text{0.36}}$
        & 19.70 & 0.704 & 0.19 $_{\scriptscriptstyle \pm \text{0.16}}$
        & 7.87 & 0.633 &  0.29 $_{\scriptscriptstyle \pm \text{0.18}}$
        & 9.73 & 0.699 & 0.32 $_{\scriptscriptstyle \pm \text{0.15}}$ \\
    \midrule
    \multicolumn{1}{l||}{\textbf{CosyVoice 2}}
        & 2.09 & 0.709 & - 
        & 8.12 & 0.696 & - 
        & 33.36 & 0.672 & - 
        & 8.78  & 0.600 & - 
        & 13.09 & 0.669 & - \\
    \multicolumn{1}{r||}{\textbf{w/ INTP}}
        & 1.65 & 0.709 & 0.24 $_{\scriptscriptstyle \pm \text{0.25}}$
        & 6.87 & 0.696 & 0.20 $_{\scriptscriptstyle \pm \text{0.16}}$
        & 28.31 & 0.671 & 0.63 $_{\scriptscriptstyle \pm \text{0.30}}$
        & 5.39  & 0.603 & 0.28 $_{\scriptscriptstyle \pm \text{0.31}}$
        & 10.56 & 0.670 & 0.33 $_{\scriptscriptstyle \pm \text{0.12}}$ \\
    \midrule
    \multicolumn{1}{l||}{\textbf{Ints}}
        & 3.14 & 0.688 & -
        & 12.08 & 0.666 & - 
        & 22.88 & 0.646 & - 
        & 9.78  & 0.572 & - 
        & 11.97 & 0.643 & - \\
    \multicolumn{1}{r||}{\textbf{w/ INTP}}
        & 2.36 & 0.686 & 0.20 $_{\scriptscriptstyle \pm \text{0.36}}$
        & 9.38  & 0.664 & 0.11 $_{\scriptscriptstyle \pm \text{0.22}}$
        & 13.80 & 0.642 & 0.20 $_{\scriptscriptstyle \pm \text{0.38}}$
        & 6.28  & 0.571 & 0.18 $_{\scriptscriptstyle \pm \text{0.23}}$
        & 7.96  & 0.641 & 0.17 $_{\scriptscriptstyle \pm \text{0.15}}$ \\
    \bottomrule
    \end{tabular}
    }
\end{threeparttable}
\caption{Improvements of DPO with INTP for different models (\textbf{AR-based}: ARS~\cite{maskgct}, CosyVoice 2~\cite{cosyvoice}, and Ints (Appendix~\ref{sec:appendix-detail-ins}). \textbf{FM-based}: F5-TTS~\cite{f5tts}. \textbf{MGM-based}: MaskGCT~\cite{maskgct}) on diverse domains. ARS, F5-TTS, and MaskGCT participated in the INTP construction, while CosyVoice 2 and Ints did not.}
\vspace{-3mm}
\label{tab:expt-main}
\end{center}
\end{table*}

\subsection{DPO for Masked Generative Models}
\label{method:dpo_mgm}
Masked generative model (MGM) is a type of Non-AR generative model, which is also widely adopted in speech generation, as seen in models such as NaturalSpeech 3~\cite{naturalspeech3}, and MaskGCT~\cite{maskgct}. MGM aims to recover a discrete sequence \(y = [z_1, z_2, \ldots, z_n]\) from its partially masked version \(y_t = y \odot m_t\), where \(m_t \in \{0,1\}^n\) is a binary mask sampled via a schedule \(\gamma(t) \in (0,1]\). MGM is trained to predict masked tokens from unmasked tokens and condition $x$, modeled as $p_\theta(y_0 \mid y_t, x)$, optimizing the sum of the marginal cross-entropy for each unmasked token:
\vspace{-2mm}
\begin{equation}
\small{
\mathcal{L}_{\text{mask}} = -\mathbb{E}_{y, x, t, m_t} \sum_{i=1}^{n} m_{t,i} \cdot \log p_{\theta}(z_i \mid y_t, x).
}
\end{equation}
Using a similar derivation as in Section~\ref{method:dpo_fm}, we extend DPO for MGM. Let \(p_{\text{ref}}(y_0 \mid y_t, x)\) represent the reference policy. The DPO loss for MGM is given by:
\vspace{-2mm}
\begin{equation}
{\scriptsize
\begin{aligned}
& \mathcal{L}_{\text{DPO-MGM}} = -\mathbb{E}_{(y^w, y^l, x) \sim \mathcal{D}, t}  \\
& \log \sigma \left( \beta \left( \log \frac{p_\theta(y^w_0|y^w_t,x)}{p_{\text{ref}}(y^w_0|y^w_t,x)}
 - \log \frac{p_\theta(y^l_0|y^l_t,x)}{p_{\text{ref}}(y^l_0|y^l_t,x)} \right) \right).
\end{aligned}
}
\end{equation}
Here, \(y^w_t\) and \(y^l_t\) are masked versions of \(y^w_0\) and \(y^l_0\). Note that $p_\theta(y_0|y_t,x)$ corresponds to the sum of the log-probabilities of the unmasked tokens in the context of MGM. We provide more details about the derivation in Appendix~\ref{sec:appendix-derivation-mgm}. In this study, we select MaskGCT~\cite{maskgct} as a representative to apply this proposed algorithm for its text-to-codec stage.

\section{Experiments}


\paragraph{Evaluation Data} 
We evaluate zero-shot TTS systems across diverse domains in both English and Chinese languages. Based on SeedTTS's evaluation samples~\cite{seedtts} (which are widely used and also serve as the evaluation set for the pre-trained models of ARS~\cite{maskgct}, F5-TTS~\cite{f5tts}, MaskGCT~\cite{maskgct}, and CosyVoice 2~\cite{cosyvoice2} in this study), we construct evaluation sets across four distinct domains: (1) \textbf{Regular cases:} We use SeedTTS test-en (1,000 samples) and SeedTTS test-zh datasets (2,000 samples). (2) \textbf{Articulatory cases:} These involve tongue twisters and repeated texts. For Chinese, we use SeedTTS test-hard, while for English, we use reference speech prompts of SeedTTS test-en, and employ Deepseek-V3~\cite{deepseek-v3} to construct the articulatory texts like SeedTTS test-hard. There are 800 samples in total. (3) \textbf{Code-switching cases:} These target texts are a mixture of English and Chinese. Based on SeedTTS test-en and test-zh, we keep their reference speech prompts unchanged, and adopt Deepseek-V3 to transform their texts into code-switching style. There are 1,000 samples in total. (4) \textbf{Cross-lingual cases:} We construct two types of cross-lingual samples: \textit{zh2en} (500 samples) and \textit{en2zh} (500 samples). The zh2en means Chinese reference speech (from SeedTTS test-zh) with English target text (from SeedTTS test-en). Similarly for en2zh. The detailed distribution of these sets is presented in Table~\ref{tab:eval-data}, Appendix~\ref{sec:appendix-eval-data}.

\paragraph{Evaluation Metrics} For objective metrics, we evaluate the intelligibility (WER, $\downarrow$), speaker similarity (SIM, $\uparrow$), and overall speech quality (UTMOS~\cite{utmos}, $\uparrow$). Specifically, for WER, we employ \texttt{Whisper-large-v3}~\cite{whisper} for English, and \texttt{Paraformer-zh}~\cite{paraformer, funasr} for Chinese and code-switching texts. For SIM, we compute the cosine similarity between the WavLM TDNN~\cite{wavlm} speaker embeddings of generated samples and the reference speeches. For subjective metrics, we employ Comparative Mean Opinion Score (rated from -2 to 2) to evaluate naturalness (N-CMOS, $\uparrow$), use reading accuracy (Section~\ref{sec:intp-human-verify}) to evaluate intelligibility, and use A/B Testing to compare speaker similarity between the generated samples before and after intelligibility alignment. Detailed descriptions of all the metrics are provided in Appendix~\ref{sec:appendix-eval}.

\begin{figure*}[t]
    \centering
    \begin{subfigure}[b]{0.35\textwidth}
        \centering
        \includegraphics[width=\textwidth]{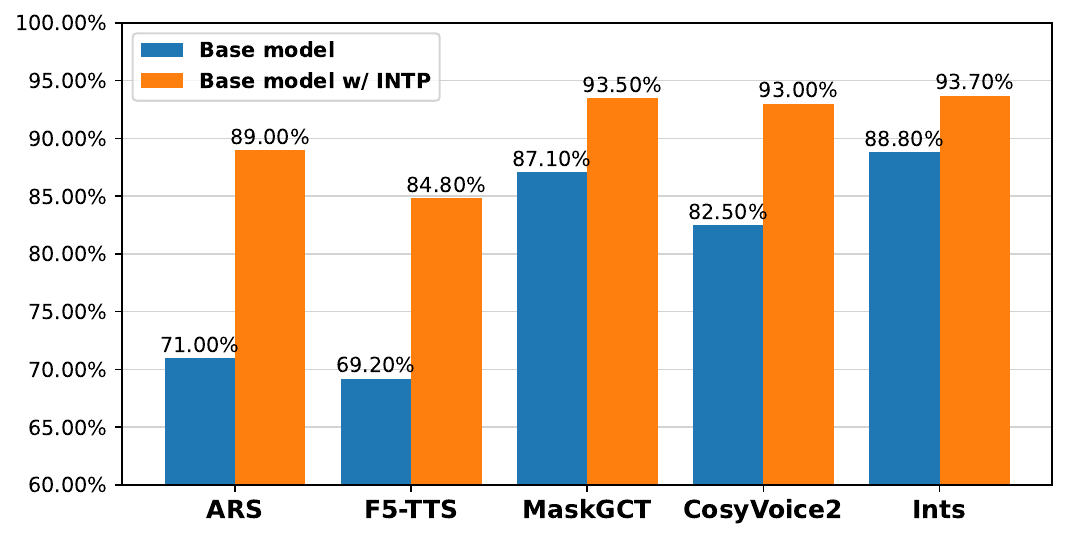}
        \caption{Comparison of reading accuracy.}
        \label{fig:sub-eval-intelligibility}
        \vspace{-2mm}
    \end{subfigure}
    \hfill
    \begin{subfigure}[b]{0.63\textwidth}
        \centering
        \includegraphics[width=\textwidth]{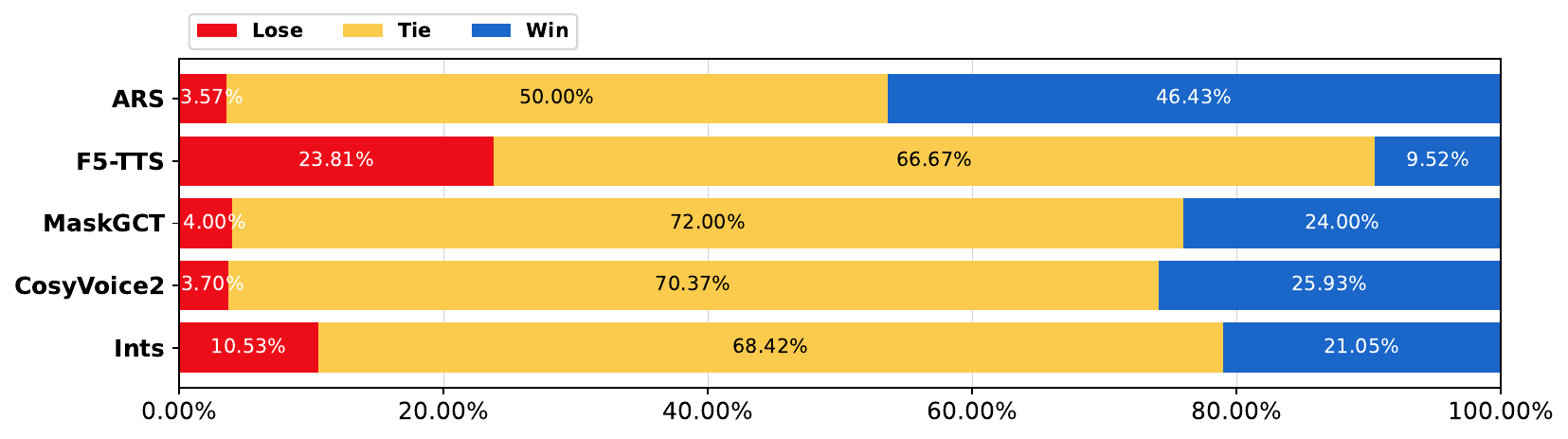}
        \caption{Win/Lose/Tie of speaker similarity after INTP alignment.}
        \label{fig:sub-eval-similarity}
        \vspace{-2mm}
    \end{subfigure}
    \caption{Subjective evaluation of intelligibility and speaker similarity for models before and after INTP alignment.}
    \label{fig:sub-eval}
    \vspace{-4mm}
\end{figure*}

\subsection{Effect of DPO with INTP}\label{sec:expt-main}

To verify the effectiveness of DPO with INTP for existing TTS models, we conduct alignment experiments with multiple models. In addition to ARS, F5-TTS, and MaskGCT, which were used in constructing the INTP dataset, we also introduce two more powerful models in terms of intelligibility: CosyVoice 2~\cite{cosyvoice2} and Ints (Appendix~\ref{sec:appendix-detail-ins}), to validate INTP's weak-to-strong generalization capability. The experimental results are presented in Table~\ref{tab:expt-main}, including results on the objective WER, SIM, and the subjective naturalness CMOS.

We observe three key findings from Table~\ref{tab:expt-main}: (1) Across different evaluation cases, while almost all models demonstrate strong intelligibility performance in regular cases (WER < 4.0), they struggle significantly with articulatory, code-switching, and cross-lingual cases. We show some hallucinated outputs for these domains on our demo website. (2) Comparing across models, CosyVoice 2 and Ints achieves better average intelligibility performance across all cases (WER of 13.09 and 11.97), highlighting the strength of using a textual LLM as the initialization of large-scale TTS model~\cite{cosyvoice2}. (3) Through DPO with INTP, all models, including the more intelligible CosyVoice 2 and Ints that are out of the INTP distribution, show improvements in both intelligibility (WER) and naturalness (N-CMOS), and display comparable performance for speaker similarity (SIM). 

Furthermore, we randomly sample 300 samples for subjective evaluation, including assessments of reading accuracy and A/B testing of speaker similarity before and after INTP alignment (see Appendix~\ref{sec:appendix-sub-eval} for details). The results in Figure~\ref{fig:sub-eval} demonstrate that INTP alignment enhances all five models in terms of both intelligibility (higher reading accuracy in Figure~\ref{fig:sub-eval-intelligibility}) and speaker similarity (more Tie/Win percentages in Figure~\ref{fig:sub-eval-similarity}).

\subsection{Effect of Different Data within INTP}

\begin{table*}[t]
\begin{center}
\resizebox{\textwidth}{!}{
\begin{threeparttable}
    \begin{tabular}{c||rrc|rrc|rrc|rrc||rrc}
    \toprule
    \multirow{2}{*}{\textbf{Model}} & 
    \multicolumn{3}{c|}{\textbf{Regular cases}} &
    \multicolumn{3}{c|}{\textbf{Articulatory cases}} &
    \multicolumn{3}{c|}{\textbf{Code-switching cases}} &
    \multicolumn{3}{c||}{\textbf{Cross-lingual cases}} &
    \multicolumn{3}{c}{\textbf{Avg}} \\
    \cmidrule(lr){2-13} \cmidrule(lr){14-16}
     & \textbf{WER} & \textbf{SIM} & \textbf{UTMOS} & \textbf{WER} & \textbf{SIM} & \textbf{UTMOS} & \textbf{WER} & \textbf{SIM} & \textbf{UTMOS} & \textbf{WER} & \textbf{SIM} & \textbf{UTMOS} & \textbf{WER} & \textbf{SIM} & \textbf{UTMOS} \\
    \midrule 
    \rowcolor{gray!30} \multicolumn{16}{c}{\textbf{\textit{Group 1: Effect of Data across Different Text Types}}} \\
    \midrule 
    \multicolumn{1}{l||}{\textbf{ARS}~\cite{maskgct}} & 3.96 & 0.717 & 3.145 & 20.03 & 0.693 & 2.915 & 54.15 & 0.693 & 3.045 & 19.76 & 0.630 & 3.120 & 24.47 & 0.683 & 3.056 \\
    \cmidrule(lr){1-16}
    \multicolumn{1}{r||}{\textbf{w/ Regular}} & 2.45 & 0.727 & 3.200 & 17.41 & 0.706 & 3.000 & 37.52 & \textbf{0.701} & 3.110 & 9.66 & \textbf{0.638} & 3.200 & 16.76 & \textbf{0.693} & 3.128 \\
    \multicolumn{1}{r||}{\textbf{w/ Repeated}} & 2.33 & 0.725 & 3.225 & 12.88 & 0.711 & 3.050 & 39.74 & \textbf{0.701} & 3.150 & 10.96 & 0.636 & 3.235 & 16.48 & \textbf{0.693} & 3.165 \\
    \multicolumn{1}{r||}{\textbf{w/ Code-switching}} & 2.32 & \textbf{0.729} & 3.220 & 17.67 & 0.704 & 3.050 & \textbf{34.20} & 0.695 & 3.140 & 8.69 & 0.633 & 3.215 & 15.72 & 0.690 & 3.156 \\
    \multicolumn{1}{r||}{\textbf{w/ Pronunciation-perturbed}} & \textbf{2.21} & 0.720 & \textbf{3.250} & 17.76 & 0.693 & \textbf{3.075} & 35.99 & 0.687 & 3.185 & \textbf{8.24} & 0.617 & \textbf{3.285} & 16.05 & 0.679 & \textbf{3.199} \\
    \multicolumn{1}{r||}{\textbf{w/ Punctuation-perturbed}} & 2.46 & 0.722 & 3.240 & 17.35 & 0.699 & 3.020 & 42.73 & 0.694 & 3.160 & 10.94 & 0.624 & 3.255 & 18.37 & 0.684 & 3.169 \\
    \midrule
    \multicolumn{1}{r||}{\textbf{w/ INTP}} & 2.32 & 0.727 & 3.210 & \textbf{12.83} & \textbf{0.713} & 3.035 & 36.91 & 0.698 & 3.145 & 9.57 & 0.632 & 3.250 & \textbf{15.41} & 0.692 & 3.160 \\
    \midrule 
    \rowcolor{gray!30} \multicolumn{16}{c}{\textbf{\textit{Group 2: Effect of Data across Different Models}}} \\
    \midrule 
    \multicolumn{1}{l||}{\textbf{ARS}~\cite{maskgct}} & 3.96 & 0.717 & 3.145 & 20.03 & 0.693 & 2.915 & 54.15 & 0.693 & 3.045 & 19.76 & 0.630 & 3.120 & 24.47 & 0.683 & 3.056 \\ \cmidrule(lr){1-16}
    \multicolumn{1}{r||}{\textbf{w/ ARS pairs}} & 2.56 & 0.717 & 3.200 & 13.05 & 0.705 & 3.015 & 40.91 & 0.691 & 3.125 & 11.07 & 0.622 & 3.225 & 16.90 & 0.684 & 3.141 \\
    \multicolumn{1}{r||}{\textbf{w/ MaskGCT pairs}} & 2.37 & 0.724 & \textbf{3.210} & 16.85 & 0.700 & 3.010 & 37.41 & 0.692 & 3.105 & \textbf{8.83} & 0.625 & 3.200 & 16.37 & 0.685 & 3.131 \\
    \multicolumn{1}{r||}{\textbf{w/ F5-TTS pairs}} & 2.46 & 0.721 & \textbf{3.210} & 14.99 & 0.705 & \textbf{3.035} & 38.77 & 0.690 & 3.115 & 10.01 & 0.621 & 3.225 & 16.56 & 0.684 & 3.146 \\ \midrule
    \multicolumn{1}{r||}{\textbf{w/ Intra pairs}} & 2.33 & 0.721 & 3.200 & 15.29 & 0.705 & 3.015 & 37.99 & 0.687 & 3.115 & 9.36 & 0.624 & 3.200 & 16.24 & 0.684 & 3.133 \\
    \multicolumn{1}{r||}{\textbf{w/ Inter pairs}} & \textbf{2.25} & 0.726 & 3.180 & 15.42 & 0.703 & 2.965 & 38.69 & 0.697 & 3.065 & 10.61 & 0.631 & 3.170 & 16.74 & 0.689 & 3.095 \\ \midrule
    \multicolumn{1}{r||}{\textbf{w/ INTP}} & 2.32 & \textbf{0.727} & \textbf{3.210} & \textbf{12.83} & \textbf{0.713} & \textbf{3.035} & \textbf{36.91} & \textbf{0.698} & \textbf{3.145} & 9.57 & \textbf{0.632} & \textbf{3.250} & \textbf{15.41} & \textbf{0.692} & \textbf{3.160} \\
    \midrule 
    \rowcolor{gray!30} \multicolumn{16}{c}{\textbf{\textit{Group 3: Effect of Different Negative Samples}}} \\
    \midrule 
    \multicolumn{1}{l||}{\textbf{ARS}~\cite{maskgct}} & 3.96 & 0.717 & 3.145 & 20.03 & 0.693 & 2.915 & 54.15 & 0.693 & 3.045 & 19.76 & 0.630 & 3.120 & 24.47 & 0.683 & 3.056 \\ \cmidrule(lr){1-16}
    \multicolumn{1}{r||}{\textbf{w/ Regular (SFT)}$^*$} & 3.28 & 0.716 & 3.165 & 20.03 & 0.685 & 2.935 & 48.73 & 0.691 & 3.065 & 17.25 & 0.630 & 3.165 & 22.32 & 0.680 & 3.083 \\
    \multicolumn{1}{r||}{\textbf{w/ Regular$^*$}} & 2.45 & \textbf{0.727} & 3.200 & 17.41 & \textbf{0.706} & 3.000 & 37.52 & \textbf{0.701} & 3.110 & 9.66 & \textbf{0.638} & 3.200 & 16.76 & \textbf{0.693} & 3.128 \\
    \multicolumn{1}{r||}{\textbf{w/ Pronunciation-perturbed}$^*$} & \textbf{2.21} & 0.720 & \textbf{3.250} & 17.76 & 0.693 & \textbf{3.075} & \textbf{35.99} & 0.687 & \textbf{3.185} & \textbf{8.24} & 0.617 & \textbf{3.285} & \textbf{16.05} & 0.679 & \textbf{3.199} \\
    \multicolumn{1}{r||}{\textbf{w/ Punctuation-perturbed}$^*$} & 2.46 & 0.722 & 3.240 & \textbf{17.35} & 0.699 & 3.020 & 42.73 & 0.694 & 3.160 & 10.94 & 0.624 & 3.255 & 18.37 & 0.684 & 3.169 \\
    \bottomrule
    \end{tabular}
    \begin{tablenotes}
        \item[*] The positive samples in these four experiments are identical. \textbf{w/ Regular (SFT)} refers to supervised fine-tuning using positive samples only, excluding negative samples. \textbf{w/ Regular} employs WER-based negative samples, while the other two utilize our proposed human-guided negative samples.
    \end{tablenotes}
\end{threeparttable}
}
\caption{Effect of different data within INTP for ARS.}
\label{tab:expt-abs-total}
\end{center}
\end{table*}

To investigate the impact of different distributions within INTP, we conduct ablation studies from multiple perspectives. In Table~\ref{tab:expt-abs-total}, we present three groups of experiments on ARS: the effect of data across different text types, across different models, and the effect of different negative samples. Additional results, including the effect of data across different languages are provided in Appendix~\ref{sec:appendix-more-results}.

We observe three key findings from Table~\ref{tab:expt-abs-total}: (1) Group 1 demonstrates that different scenarios require customized post-training data. For instance, repeated data proves particularly effective for articulatory cases, while pronunciation-perturbed data significantly improves pronunciation accuracy and WER in cross-lingual cases (see our demo website for details). Moreover, utilizing data from multiple scenarios (i.e., the complete INTP) yields the best overall improvements. (2) Group 2 reveals that model improvement can be achieved through alignment using synthetic data, regardless of whether it's generated by the model itself or other models. Besides, the intra-pairs and inter-pairs are complementary for model improvements. (3) Group 3 shows that using only positive samples from INTP for supervised fine-tuning (SFT) can already improve quality. Building upon this, incorporating negative samples for preference learning leads to even more substantial gains. 


\subsection{Iterative Intelligibility Alignment}\label{sec:iterative-align}

\begin{table*}[t]
\vspace{-3mm}
\begin{center}
\begin{threeparttable}
    \resizebox{\textwidth}{!}{
    \begin{tabular}{c|c||rcc|rcc|rcc|rcc||rcc}
    \toprule
    \multirow{2}{*}{\textbf{Model}} & 
    \multirow{2}{*}{\textbf{Preference Data}} & 
    \multicolumn{3}{c|}{\textbf{Regular cases}} &
    \multicolumn{3}{c|}{\textbf{Articulatory cases}} & 
    \multicolumn{3}{c|}{\textbf{Code-switching cases}} &
    \multicolumn{3}{c||}{\textbf{Cross-lingual cases}} &
    \multicolumn{3}{c}{\textbf{Avg}} \\
    \cmidrule(lr){3-14} \cmidrule(lr){15-17}
     & & \textbf{WER} & \textbf{SIM} & \textbf{UTMOS} & \textbf{WER} & \textbf{SIM} & \textbf{UTMOS} & \textbf{WER} & \textbf{SIM} & \textbf{UTMOS} & \textbf{WER} & \textbf{SIM} & \textbf{UTMOS} & \textbf{WER} & \textbf{SIM} & \textbf{UTMOS} \\
    \midrule
    \textbf{Ints} & - 
        & 3.14 & 0.688 & 3.175 
        & 12.08 & 0.666 & 3.025 
        & 22.88 & 0.646 & 3.045 
        & 9.78 & 0.572 & 3.150 
        & 11.97 & 0.643 & 3.099 \\
    \textbf{Ints v1} & \textbf{INTP} 
        & 2.36 & 0.686 & 3.205 
        & 9.38 & 0.664 & 3.060 
        & 13.80 & 0.642 & 3.125 
        & 6.28 & 0.571 & 3.230 
        & 7.96 & 0.641 & 3.155 \\
    \textbf{Ints v2} & \textbf{Ints v1 generated} 
        & 2.21 & 0.686 & 3.210 
        & 8.48 & 0.660 & 3.085 
        & 12.33 & 0.643 & 3.140 
        & 5.40 & 0.567 & 3.250 
        & 7.10 & 0.639 & 3.171 \\
    \bottomrule
    \end{tabular}
    }
\end{threeparttable}
\caption{Iterative Preference Alignment for Ints.}
\label{tab:data_flywheel}
\end{center}
\vspace{-3mm}
\end{table*}


Furthermore, we explore how to establish an \textit{iterative} preference alignment, i.e., data and model flywheel~\cite{RLHF-anthoropic,llama3,iterative-dpo}. This approach aligns with the online reinforcement learning (RL) framework \citet{li2023rema4}.  We investigate two rounds of alignment based on Ints, where Ints v1 (INTP-aligned model) is used to generate new preference data for training Ints v2, following a similar \textit{cadence} of data collection as~\cite{RLHF-anthoropic}. To prepare Ints v1 generated preference data, we sample a challenging prompt subset from INTP and adopt the same pipeline as INTP to construct preference pairs (see Appendix~\ref{sec:appendix-ints-training-data} for details). The results of this iterative alignment are shown in Table~\ref{tab:data_flywheel}. We can observe that compared to Ints v1, Ints v2 yields additional improvements across all scenarios, which demonstrates that effectiveness of iterative alignment. However, we observe that the magnitude of improvement in the second round is notably smaller than the first round. We suspect this indicates that the upper bound of iterative alignment is largely determined by the base model's inherent capabilities, suggesting future research should focus on base models with higher potential.

\section{Conclusion}

In this work, we focus on the intelligibility issues of modern zero-shot TTS systems across diverse domains, especially in hard-to-pronounce texts, code-switching, and cross-lingual synthesis. We propose to address these challenges using preference alignment with our newly constructed INTP dataset, which contains diverse preference pairs determined through model self-comparison, cross-model comparison, and human guidance. We employ DPO and design  special extensions to significantly improve various TTS architectures, while demonstrating INTP's weak-to-strong generalization capability and establishing an iterative preference alignment flywheel with more powerful base models.


\section*{Limitations}

While our approach demonstrates significant improvements in zero-shot TTS intelligibility across diverse domains, several limitations remain. Although INTP covers multiple challenging scenarios, it may not fully capture all edge cases, such as specialized jargon or rare language pairs. Future work could expand to more low-resource languages and niche domains. Besides, constructing INTP and conduct alignment experiments on large models like Ints require substantial computational resources, potentially limiting accessibility.

\section*{Potential Risks}

The proposed method introduces several risks that warrant consideration. Enhanced TTS systems could be exploited to generate deceptive content (e.g., deepfake audio), posing ethical challenges. Robust safeguards and watermarking mechanisms are critical for deployment. While INTP uses public datasets, real-world applications may risk incorporating sensitive or copyrighted speech data, requiring strict governance protocols.

\section*{Acknowledgment}
This work is partially supported by the NSFC under Grant 62376237, Shenzhen Science and Technology Program ZDSYS20230626091302006, and Shenzhen Research Institute of Big Data (Internal Project Fund, Grant No. T00120230002). We appreciate Yushun Zhang and the anonymous reviewers for their insightful comments and suggestions.

\bibliography{custom}

\begin{thebibliography}{68}
\providecommand{\natexlab}[1]{#1}

\bibitem[{Abdin et~al.(2024)Abdin, Jacobs, Awan, Aneja, Awadallah, Awadalla, Bach, Bahree, Bakhtiari, Behl, Benhaim, Bilenko, Bjorck, Bubeck, Cai, Mendes, Chen, Chaudhary, Chopra, Giorno, de~Rosa, Dixon, Eldan, Iter, Garg, Goswami, Gunasekar, Haider, Hao, Hewett, Huynh, Javaheripi, Jin, Kauffmann, Karampatziakis, Kim, Khademi, Kurilenko, Lee, Lee, Li, Liang, Liu, Lin, Lin, Madan, Mitra, Modi, Nguyen, Norick, Patra, Perez{-}Becker, Portet, Pryzant, Qin, Radmilac, Rosset, Roy, Ruwase, Saarikivi, Saied, Salim, Santacroce, Shah, Shang, Sharma, Song, Tanaka, Wang, Ward, Wang, Witte, Wyatt, Xu, Xu, Yadav, Yang, Yang, Yu, Zhang, Zhang, Zhang, Zhang, Zhang, Zhang, Zhang, and Zhou}]{phi3}
Marah~I Abdin, Sam~Ade Jacobs, Ammar~Ahmad Awan, Jyoti Aneja, Ahmed Awadallah, Hany Awadalla, Nguyen Bach, Amit Bahree, Arash Bakhtiari, Harkirat~S. Behl, Alon Benhaim, Misha Bilenko, Johan Bjorck, S{\'{e}}bastien Bubeck, Martin Cai, Caio C{\'{e}}sar~Teodoro Mendes, Weizhu Chen, Vishrav Chaudhary, Parul Chopra, Allie~Del Giorno, Gustavo de~Rosa, Matthew Dixon, Ronen Eldan, Dan Iter, Amit Garg, Abhishek Goswami, Suriya Gunasekar, Emman Haider, Junheng Hao, Russell~J. Hewett, Jamie Huynh, Mojan Javaheripi, Xin Jin, Piero Kauffmann, Nikos Karampatziakis, Dongwoo Kim, Mahoud Khademi, Lev Kurilenko, James~R. Lee, Yin~Tat Lee, Yuanzhi Li, Chen Liang, Weishung Liu, Eric Lin, Zeqi Lin, Piyush Madan, Arindam Mitra, Hardik Modi, Anh Nguyen, Brandon Norick, Barun Patra, Daniel Perez{-}Becker, Thomas Portet, Reid Pryzant, Heyang Qin, Marko Radmilac, Corby Rosset, Sambudha Roy, Olatunji Ruwase, Olli Saarikivi, Amin Saied, Adil Salim, Michael Santacroce, Shital Shah, Ning Shang, Hiteshi Sharma, Xia Song, Masahiro Tanaka,
  Xin Wang, Rachel Ward, Guanhua Wang, Philipp Witte, Michael Wyatt, Can Xu, Jiahang Xu, Sonali Yadav, Fan Yang, Ziyi Yang, Donghan Yu, Chengruidong Zhang, Cyril Zhang, Jianwen Zhang, Li~Lyna Zhang, Yi~Zhang, Yue Zhang, Yunan Zhang, and Xiren Zhou. 2024.
\newblock Phi-3 technical report: {A} highly capable language model locally on your phone.
\newblock \emph{arXiv preprint}, abs/2404.14219.

\bibitem[{Anastassiou et~al.(2024)Anastassiou, Chen, Chen, Chen, Chen, Chen, Cong, Deng, Ding, Gao, Gong, Huang, Huang, Huang, Huo, Jia, Li, Li, Li, Li, Li, Li, Liu, Liu, Liu, Liu, Liu, Liu, Lu, Pan, Wang, Wang, Wang, Wei, Wu, Yao, Yang, Yi, Zhang, Zhang, Zhang, Zhang, Zhang, Zhao, Zhong, and Zhuang}]{seedtts}
Philip Anastassiou, Jiawei Chen, Jitong Chen, Yuanzhe Chen, Zhuo Chen, Ziyi Chen, Jian Cong, Lelai Deng, Chuang Ding, Lu~Gao, Mingqing Gong, Peisong Huang, Qingqing Huang, Zhiying Huang, Yuanyuan Huo, Dongya Jia, Chumin Li, Feiya Li, Hui Li, Jiaxin Li, Xiaoyang Li, Xingxing Li, Lin Liu, Shouda Liu, Sichao Liu, Xudong Liu, Yuchen Liu, Zhengxi Liu, Lu~Lu, Junjie Pan, Xin Wang, Yuping Wang, Yuxuan Wang, Zhen Wei, Jian Wu, Chao Yao, Yifeng Yang, Yuanhao Yi, Junteng Zhang, Qidi Zhang, Shuo Zhang, Wenjie Zhang, Yang Zhang, Zilin Zhao, Dejian Zhong, and Xiaobin Zhuang. 2024.
\newblock Seed-tts: {A} family of high-quality versatile speech generation models.
\newblock \emph{arXiv preprint}, abs/2406.02430.

\bibitem[{Ardila et~al.(2019)Ardila, Branson, Davis, Henretty, Kohler, Meyer, Morais, Saunders, Tyers, and Weber}]{ardila2019common}
Rosana Ardila, Megan Branson, Kelly Davis, Michael Henretty, Michael Kohler, Josh Meyer, Reuben Morais, Lindsay Saunders, Francis~M Tyers, and Gregor Weber. 2019.
\newblock Common voice: A massively-multilingual speech corpus.
\newblock \emph{arXiv preprint arXiv:1912.06670}.

\bibitem[{Bai et~al.(2022)Bai, Jones, Ndousse, Askell, Chen, DasSarma, Drain, Fort, Ganguli, Henighan, Joseph, Kadavath, Kernion, Conerly, Showk, Elhage, Hatfield{-}Dodds, Hernandez, Hume, Johnston, Kravec, Lovitt, Nanda, Olsson, Amodei, Brown, Clark, McCandlish, Olah, Mann, and Kaplan}]{RLHF-anthoropic}
Yuntao Bai, Andy Jones, Kamal Ndousse, Amanda Askell, Anna Chen, Nova DasSarma, Dawn Drain, Stanislav Fort, Deep Ganguli, Tom Henighan, Nicholas Joseph, Saurav Kadavath, Jackson Kernion, Tom Conerly, Sheer~El Showk, Nelson Elhage, Zac Hatfield{-}Dodds, Danny Hernandez, Tristan Hume, Scott Johnston, Shauna Kravec, Liane Lovitt, Neel Nanda, Catherine Olsson, Dario Amodei, Tom~B. Brown, Jack Clark, Sam McCandlish, Chris Olah, Benjamin Mann, and Jared Kaplan. 2022.
\newblock Training a helpful and harmless assistant with reinforcement learning from human feedback.
\newblock \emph{arXiv preprint}, abs/2204.05862.

\bibitem[{Borsos et~al.(2023{\natexlab{a}})Borsos, Marinier, Vincent, Kharitonov, Pietquin, Sharifi, Roblek, Teboul, Grangier, Tagliasacchi, and Zeghidour}]{audiolm}
Zal{\'{a}}n Borsos, Rapha{\"{e}}l Marinier, Damien Vincent, Eugene Kharitonov, Olivier Pietquin, Matthew Sharifi, Dominik Roblek, Olivier Teboul, David Grangier, Marco Tagliasacchi, and Neil Zeghidour. 2023{\natexlab{a}}.
\newblock Audiolm: {A} language modeling approach to audio generation.
\newblock \emph{{IEEE} {ACM} Trans. Audio Speech Lang. Process.}, 31:2523--2533.

\bibitem[{Borsos et~al.(2023{\natexlab{b}})Borsos, Sharifi, Vincent, Kharitonov, Zeghidour, and Tagliasacchi}]{soundstorm}
Zal{\'{a}}n Borsos, Matthew Sharifi, Damien Vincent, Eugene Kharitonov, Neil Zeghidour, and Marco Tagliasacchi. 2023{\natexlab{b}}.
\newblock Soundstorm: Efficient parallel audio generation.
\newblock \emph{arXiv preprint}, abs/2305.09636.

\bibitem[{Burns et~al.(2024)Burns, Izmailov, Kirchner, Baker, Gao, Aschenbrenner, Chen, Ecoffet, Joglekar, Leike, Sutskever, and Wu}]{weak-to-strong-generalization}
Collin Burns, Pavel Izmailov, Jan~Hendrik Kirchner, Bowen Baker, Leo Gao, Leopold Aschenbrenner, Yining Chen, Adrien Ecoffet, Manas Joglekar, Jan Leike, Ilya Sutskever, and Jeffrey Wu. 2024.
\newblock Weak-to-strong generalization: Eliciting strong capabilities with weak supervision.
\newblock In \emph{{ICML}}. OpenReview.net.

\bibitem[{Chen et~al.(2024{\natexlab{a}})Chen, Hu, Wu, Wang, Chng, and Zhang}]{uno}
Chen Chen, Yuchen Hu, Wen Wu, Helin Wang, Eng~Siong Chng, and Chao Zhang. 2024{\natexlab{a}}.
\newblock Enhancing zero-shot text-to-speech synthesis with human feedback.
\newblock \emph{arXiv preprint}, abs/2406.00654.

\bibitem[{Chen et~al.(2024{\natexlab{b}})Chen, Byun, Elsner, and Perrault}]{dlpo}
Jingyi Chen, Ju-Seung Byun, Micha Elsner, and Andrew Perrault. 2024{\natexlab{b}}.
\newblock Dlpo: Diffusion model loss-guided reinforcement learning for fine-tuning text-to-speech diffusion models.
\newblock \emph{arXiv preprint}, abs/2405.14632.

\bibitem[{Chen et~al.(2022)Chen, Wang, Chen, Wu, Liu, Chen, Li, Kanda, Yoshioka, Xiao et~al.}]{wavlm}
Sanyuan Chen, Chengyi Wang, Zhengyang Chen, Yu~Wu, Shujie Liu, Zhuo Chen, Jinyu Li, Naoyuki Kanda, Takuya Yoshioka, Xiong Xiao, et~al. 2022.
\newblock Wavlm: Large-scale self-supervised pre-training for full stack speech processing.
\newblock \emph{IEEE Journal of Selected Topics in Signal Processing}, 16(6):1505--1518.

\bibitem[{Chen et~al.(2024{\natexlab{c}})Chen, Niu, Ma, Deng, Wang, Zhao, Yu, and Chen}]{f5tts}
Yushen Chen, Zhikang Niu, Ziyang Ma, Keqi Deng, Chunhui Wang, Jian Zhao, Kai Yu, and Xie Chen. 2024{\natexlab{c}}.
\newblock {F5-TTS:} {A} fairytaler that fakes fluent and faithful speech with flow matching.
\newblock \emph{arXiv preprint}, abs/2410.06885.

\bibitem[{Cideron et~al.(2024)Cideron, Girgin, Verzetti, Vincent, Kastelic, Borsos, McWilliams, Ungureanu, Bachem, Pietquin, Geist, Hussenot, Zeghidour, and Agostinelli}]{musicrl}
Geoffrey Cideron, Sertan Girgin, Mauro Verzetti, Damien Vincent, Matej Kastelic, Zal{\'{a}}n Borsos, Brian McWilliams, Victor Ungureanu, Olivier Bachem, Olivier Pietquin, Matthieu Geist, L{\'{e}}onard Hussenot, Neil Zeghidour, and Andrea Agostinelli. 2024.
\newblock Musicrl: Aligning music generation to human preferences.
\newblock In \emph{{ICML}}. OpenReview.net.

\bibitem[{Dao et~al.(2022)Dao, Fu, Ermon, Rudra, and R{\'e}}]{flashattention}
Tri Dao, Dan Fu, Stefano Ermon, Atri Rudra, and Christopher R{\'e}. 2022.
\newblock Flashattention: Fast and memory-efficient exact attention with io-awareness.
\newblock \emph{Advances in Neural Information Processing Systems}, 35:16344--16359.

\bibitem[{DeepSeek-AI et~al.(2025)DeepSeek-AI, Guo, Yang, Zhang, Song, Zhang, Xu, Zhu, Ma, Wang, Bi, Zhang, Yu, Wu, Wu, Gou, Shao, Li, Gao, Liu, Xue, Wang, Wu, Feng, Lu, Zhao, Deng, Zhang, Ruan, Dai, Chen, Ji, Li, Lin, Dai, Luo, Hao, Chen, Li, Zhang, Bao, Xu, Wang, Ding, Xin, Gao, Qu, Li, Guo, Li, Wang, Chen, Yuan, Qiu, Li, Cai, Ni, Liang, Chen, Dong, Hu, Gao, Guan, Huang, Yu, Wang, Zhang, Zhao, Wang, Zhang, Xu, Xia, Zhang, Zhang, Tang, Li, Wang, Li, Tian, Huang, Zhang, Wang, Chen, Du, Ge, Zhang, Pan, Wang, Chen, Jin, Chen, Lu, Zhou, Chen, Ye, Wang, Yu, Zhou, Pan, Li, Zhou, Wu, Ye, Yun, Pei, Sun, Wang, Zeng, Zhao, Liu, Liang, Gao, Yu, Zhang, Xiao, An, Liu, Wang, Chen, Nie, Cheng, Liu, Xie, Liu, Yang, Li, Su, Lin, Li, Jin, Shen, Chen, Sun, Wang, Song, Zhou, Wang, Shan, Li, Wang, Wei, Zhang, Xu, Li, Zhao, Sun, Wang, Yu, Zhang, Shi, Xiong, He, Piao, Wang, Tan, Ma, Liu, Guo, Ou, Wang, Gong, Zou, He, Xiong, Luo, You, Liu, Zhou, Zhu, Xu, Huang, Li, Zheng, Zhu, Ma, Tang, Zha, Yan, Ren, Ren, Sha, Fu, Xu, Xie, Zhang,
  Hao, Ma, Yan, Wu, Gu, Zhu, Liu, Li, Xie, Song, Pan, Huang, Xu, Zhang, and Zhang}]{deepseek-r1}
DeepSeek-AI, Daya Guo, Dejian Yang, Haowei Zhang, Junxiao Song, Ruoyu Zhang, Runxin Xu, Qihao Zhu, Shirong Ma, Peiyi Wang, Xiao Bi, Xiaokang Zhang, Xingkai Yu, Yu~Wu, Z.~F. Wu, Zhibin Gou, Zhihong Shao, Zhuoshu Li, Ziyi Gao, Aixin Liu, Bing Xue, Bingxuan Wang, Bochao Wu, Bei Feng, Chengda Lu, Chenggang Zhao, Chengqi Deng, Chenyu Zhang, Chong Ruan, Damai Dai, Deli Chen, Dongjie Ji, Erhang Li, Fangyun Lin, Fucong Dai, Fuli Luo, Guangbo Hao, Guanting Chen, Guowei Li, H.~Zhang, Han Bao, Hanwei Xu, Haocheng Wang, Honghui Ding, Huajian Xin, Huazuo Gao, Hui Qu, Hui Li, Jianzhong Guo, Jiashi Li, Jiawei Wang, Jingchang Chen, Jingyang Yuan, Junjie Qiu, Junlong Li, J.~L. Cai, Jiaqi Ni, Jian Liang, Jin Chen, Kai Dong, Kai Hu, Kaige Gao, Kang Guan, Kexin Huang, Kuai Yu, Lean Wang, Lecong Zhang, Liang Zhao, Litong Wang, Liyue Zhang, Lei Xu, Leyi Xia, Mingchuan Zhang, Minghua Zhang, Minghui Tang, Meng Li, Miaojun Wang, Mingming Li, Ning Tian, Panpan Huang, Peng Zhang, Qiancheng Wang, Qinyu Chen, Qiushi Du, Ruiqi Ge, Ruisong
  Zhang, Ruizhe Pan, Runji Wang, R.~J. Chen, R.~L. Jin, Ruyi Chen, Shanghao Lu, Shangyan Zhou, Shanhuang Chen, Shengfeng Ye, Shiyu Wang, Shuiping Yu, Shunfeng Zhou, Shuting Pan, S.~S. Li, Shuang Zhou, Shaoqing Wu, Shengfeng Ye, Tao Yun, Tian Pei, Tianyu Sun, T.~Wang, Wangding Zeng, Wanjia Zhao, Wen Liu, Wenfeng Liang, Wenjun Gao, Wenqin Yu, Wentao Zhang, W.~L. Xiao, Wei An, Xiaodong Liu, Xiaohan Wang, Xiaokang Chen, Xiaotao Nie, Xin Cheng, Xin Liu, Xin Xie, Xingchao Liu, Xinyu Yang, Xinyuan Li, Xuecheng Su, Xuheng Lin, X.~Q. Li, Xiangyue Jin, Xiaojin Shen, Xiaosha Chen, Xiaowen Sun, Xiaoxiang Wang, Xinnan Song, Xinyi Zhou, Xianzu Wang, Xinxia Shan, Y.~K. Li, Y.~Q. Wang, Y.~X. Wei, Yang Zhang, Yanhong Xu, Yao Li, Yao Zhao, Yaofeng Sun, Yaohui Wang, Yi~Yu, Yichao Zhang, Yifan Shi, Yiliang Xiong, Ying He, Yishi Piao, Yisong Wang, Yixuan Tan, Yiyang Ma, Yiyuan Liu, Yongqiang Guo, Yuan Ou, Yuduan Wang, Yue Gong, Yuheng Zou, Yujia He, Yunfan Xiong, Yuxiang Luo, Yuxiang You, Yuxuan Liu, Yuyang Zhou, Y.~X. Zhu,
  Yanhong Xu, Yanping Huang, Yaohui Li, Yi~Zheng, Yuchen Zhu, Yunxian Ma, Ying Tang, Yukun Zha, Yuting Yan, Z.~Z. Ren, Zehui Ren, Zhangli Sha, Zhe Fu, Zhean Xu, Zhenda Xie, Zhengyan Zhang, Zhewen Hao, Zhicheng Ma, Zhigang Yan, Zhiyu Wu, Zihui Gu, Zijia Zhu, Zijun Liu, Zilin Li, Ziwei Xie, Ziyang Song, Zizheng Pan, Zhen Huang, Zhipeng Xu, Zhongyu Zhang, and Zhen Zhang. 2025.
\newblock Deepseek-r1: Incentivizing reasoning capability in llms via reinforcement learning.

\bibitem[{DeepSeek{-}AI et~al.(2024)DeepSeek{-}AI, Liu, Feng, Xue, Wang, Wu, Lu, Zhao, Deng, Zhang, Ruan, Dai, Guo, Yang, Chen, Ji, Li, Lin, Dai, Luo, Hao, Chen, Li, Zhang, Bao, Xu, Wang, Zhang, Ding, Xin, Gao, Li, Qu, Cai, Liang, Guo, Ni, Li, Wang, Chen, Chen, Yuan, Qiu, Li, Song, Dong, Hu, Gao, Guan, Huang, Yu, Wang, Zhang, Xu, Xia, Zhao, Wang, Zhang, Li, Wang, Zhang, Zhang, Tang, Li, Tian, Huang, Wang, Zhang, Wang, Zhu, Chen, Du, Chen, Jin, Ge, Zhang, Pan, Wang, Xu, Zhang, Chen, Li, Lu, Zhou, Chen, Wu, Ye, Ye, Ma, Wang, Zhou, Yu, Zhou, Pan, Wang, Yun, Pei, Sun, Xiao, and Zeng}]{deepseek-v3}
DeepSeek{-}AI, Aixin Liu, Bei Feng, Bing Xue, Bingxuan Wang, Bochao Wu, Chengda Lu, Chenggang Zhao, Chengqi Deng, Chenyu Zhang, Chong Ruan, Damai Dai, Daya Guo, Dejian Yang, Deli Chen, Dongjie Ji, Erhang Li, Fangyun Lin, Fucong Dai, Fuli Luo, Guangbo Hao, Guanting Chen, Guowei Li, H.~Zhang, Han Bao, Hanwei Xu, Haocheng Wang, Haowei Zhang, Honghui Ding, Huajian Xin, Huazuo Gao, Hui Li, Hui Qu, J.~L. Cai, Jian Liang, Jianzhong Guo, Jiaqi Ni, Jiashi Li, Jiawei Wang, Jin Chen, Jingchang Chen, Jingyang Yuan, Junjie Qiu, Junlong Li, Junxiao Song, Kai Dong, Kai Hu, Kaige Gao, Kang Guan, Kexin Huang, Kuai Yu, Lean Wang, Lecong Zhang, Lei Xu, Leyi Xia, Liang Zhao, Litong Wang, Liyue Zhang, Meng Li, Miaojun Wang, Mingchuan Zhang, Minghua Zhang, Minghui Tang, Mingming Li, Ning Tian, Panpan Huang, Peiyi Wang, Peng Zhang, Qiancheng Wang, Qihao Zhu, Qinyu Chen, Qiushi Du, R.~J. Chen, R.~L. Jin, Ruiqi Ge, Ruisong Zhang, Ruizhe Pan, Runji Wang, Runxin Xu, Ruoyu Zhang, Ruyi Chen, S.~S. Li, Shanghao Lu, Shangyan Zhou,
  Shanhuang Chen, Shaoqing Wu, Shengfeng Ye, Shengfeng Ye, Shirong Ma, Shiyu Wang, Shuang Zhou, Shuiping Yu, Shunfeng Zhou, Shuting Pan, T.~Wang, Tao Yun, Tian Pei, Tianyu Sun, W.~L. Xiao, and Wangding Zeng. 2024.
\newblock Deepseek-v3 technical report.
\newblock \emph{arXiv preprint}, abs/2412.19437.

\bibitem[{D{\'{e}}fossez et~al.(2023)D{\'{e}}fossez, Copet, Synnaeve, and Adi}]{encodec}
Alexandre D{\'{e}}fossez, Jade Copet, Gabriel Synnaeve, and Yossi Adi. 2023.
\newblock High fidelity neural audio compression.
\newblock \emph{Trans. Mach. Learn. Res.}, 2023.

\bibitem[{Du et~al.(2024{\natexlab{a}})Du, Chen, Zhang, Hu, Lu, Yang, Hu, Zheng, Gu, Ma, Gao, and Yan}]{cosyvoice}
Zhihao Du, Qian Chen, Shiliang Zhang, Kai Hu, Heng Lu, Yexin Yang, Hangrui Hu, Siqi Zheng, Yue Gu, Ziyang Ma, Zhifu Gao, and Zhijie Yan. 2024{\natexlab{a}}.
\newblock Cosyvoice: {A} scalable multilingual zero-shot text-to-speech synthesizer based on supervised semantic tokens.
\newblock \emph{arXiv preprint}, abs/2407.05407.

\bibitem[{Du et~al.(2024{\natexlab{b}})Du, Wang, Chen, Shi, Lv, Zhao, Gao, Yang, Gao, Wang, Yu, Liu, Sheng, Gu, Deng, Wang, Zhang, Yan, and Zhou}]{cosyvoice2}
Zhihao Du, Yuxuan Wang, Qian Chen, Xian Shi, Xiang Lv, Tianyu Zhao, Zhifu Gao, Yexin Yang, Changfeng Gao, Hui Wang, Fan Yu, Huadai Liu, Zhengyan Sheng, Yue Gu, Chong Deng, Wen Wang, Shiliang Zhang, Zhijie Yan, and Jingren Zhou. 2024{\natexlab{b}}.
\newblock Cosyvoice 2: Scalable streaming speech synthesis with large language models.
\newblock \emph{arXiv preprint}, abs/2412.10117.

\bibitem[{Dubey et~al.(2024)Dubey, Jauhri, Pandey, Kadian, Al{-}Dahle, Letman, Mathur, Schelten, Yang, Fan, Goyal, Hartshorn, Yang, Mitra, Sravankumar, Korenev, Hinsvark, Rao, Zhang, Rodriguez, Gregerson, Spataru, Rozi{\`{e}}re, Biron, Tang, Chern, Caucheteux, Nayak, Bi, Marra, McConnell, Keller, Touret, Wu, Wong, Ferrer, Nikolaidis, Allonsius, Song, Pintz, Livshits, Esiobu, Choudhary, Mahajan, Garcia{-}Olano, Perino, Hupkes, Lakomkin, AlBadawy, Lobanova, Dinan, Smith, Radenovic, Zhang, Synnaeve, Lee, Anderson, Nail, Mialon, Pang, Cucurell, Nguyen, Korevaar, Xu, Touvron, Zarov, Ibarra, Kloumann, Misra, Evtimov, Copet, Lee, Geffert, Vranes, Park, Mahadeokar, Shah, van~der Linde, Billock, Hong, Lee, Fu, Chi, Huang, Liu, Wang, Yu, Bitton, Spisak, Park, Rocca, Johnstun, Saxe, Jia, Alwala, Upasani, Plawiak, Li, Heafield, Stone, and et~al.}]{llama3}
Abhimanyu Dubey, Abhinav Jauhri, Abhinav Pandey, Abhishek Kadian, Ahmad Al{-}Dahle, Aiesha Letman, Akhil Mathur, Alan Schelten, Amy Yang, Angela Fan, Anirudh Goyal, Anthony Hartshorn, Aobo Yang, Archi Mitra, Archie Sravankumar, Artem Korenev, Arthur Hinsvark, Arun Rao, Aston Zhang, Aur{\'{e}}lien Rodriguez, Austen Gregerson, Ava Spataru, Baptiste Rozi{\`{e}}re, Bethany Biron, Binh Tang, Bobbie Chern, Charlotte Caucheteux, Chaya Nayak, Chloe Bi, Chris Marra, Chris McConnell, Christian Keller, Christophe Touret, Chunyang Wu, Corinne Wong, Cristian~Canton Ferrer, Cyrus Nikolaidis, Damien Allonsius, Daniel Song, Danielle Pintz, Danny Livshits, David Esiobu, Dhruv Choudhary, Dhruv Mahajan, Diego Garcia{-}Olano, Diego Perino, Dieuwke Hupkes, Egor Lakomkin, Ehab AlBadawy, Elina Lobanova, Emily Dinan, Eric~Michael Smith, Filip Radenovic, Frank Zhang, Gabriel Synnaeve, Gabrielle Lee, Georgia~Lewis Anderson, Graeme Nail, Gr{\'{e}}goire Mialon, Guan Pang, Guillem Cucurell, Hailey Nguyen, Hannah Korevaar, Hu~Xu, Hugo
  Touvron, Iliyan Zarov, Imanol~Arrieta Ibarra, Isabel~M. Kloumann, Ishan Misra, Ivan Evtimov, Jade Copet, Jaewon Lee, Jan Geffert, Jana Vranes, Jason Park, Jay Mahadeokar, Jeet Shah, Jelmer van~der Linde, Jennifer Billock, Jenny Hong, Jenya Lee, Jeremy Fu, Jianfeng Chi, Jianyu Huang, Jiawen Liu, Jie Wang, Jiecao Yu, Joanna Bitton, Joe Spisak, Jongsoo Park, Joseph Rocca, Joshua Johnstun, Joshua Saxe, Junteng Jia, Kalyan~Vasuden Alwala, Kartikeya Upasani, Kate Plawiak, Ke~Li, Kenneth Heafield, Kevin Stone, and et~al. 2024.
\newblock The llama 3 herd of models.
\newblock \emph{arXiv preprint}, abs/2407.21783.

\bibitem[{Eskimez et~al.(2024)Eskimez, Wang, Thakker, Li, Tsai, Xiao, Yang, Zhu, Tang, Tan, Liu, Zhao, and Kanda}]{e2tts}
Sefik~Emre Eskimez, Xiaofei Wang, Manthan Thakker, Canrun Li, Chung{-}Hsien Tsai, Zhen Xiao, Hemin Yang, Zirun Zhu, Min Tang, Xu~Tan, Yanqing Liu, Sheng Zhao, and Naoyuki Kanda. 2024.
\newblock {E2} {TTS:} embarrassingly easy fully non-autoregressive zero-shot {TTS}.
\newblock In \emph{{SLT}}. {IEEE}.

\bibitem[{Fu et~al.(2024)Fu, Xiao, Wang, Zhu, Zhang, Pang, Jia, and Chen}]{tldr-perturb-vision}
Deqing Fu, Tong Xiao, Rui Wang, Wang Zhu, Pengchuan Zhang, Guan Pang, Robin Jia, and Lawrence Chen. 2024.
\newblock {TLDR:} token-level detective reward model for large vision language models.
\newblock \emph{arXiv preprint}, abs/2410.04734.

\bibitem[{Gao et~al.(2024)Gao, Zhang, Chen, Zhang, and Chen}]{emo-dpo}
Xiaoxue Gao, Chen Zhang, Yiming Chen, Huayun Zhang, and Nancy~F. Chen. 2024.
\newblock Emo-dpo: Controllable emotional speech synthesis through direct preference optimization.
\newblock \emph{arXiv preprint}, abs/2409.10157.

\bibitem[{Gao et~al.(2023)Gao, Li, Wang, Luo, Shi, Chen, Li, Zuo, Du, and Zhang}]{funasr}
Zhifu Gao, Zerui Li, Jiaming Wang, Haoneng Luo, Xian Shi, Mengzhe Chen, Yabin Li, Lingyun Zuo, Zhihao Du, and Shiliang Zhang. 2023.
\newblock Funasr: {A} fundamental end-to-end speech recognition toolkit.
\newblock In \emph{{INTERSPEECH}}, pages 1593--1597. {ISCA}.

\bibitem[{Gao et~al.(2022)Gao, Zhang, McLoughlin, and Yan}]{paraformer}
Zhifu Gao, Shiliang Zhang, Ian McLoughlin, and Zhijie Yan. 2022.
\newblock Paraformer: Fast and accurate parallel transformer for non-autoregressive end-to-end speech recognition.
\newblock In \emph{{INTERSPEECH}}, pages 2063--2067. {ISCA}.

\bibitem[{Guo et~al.(2024)Guo, Liu, Shen, Wu, Xie, Xie, and Xu}]{fireredtts}
Hao-Han Guo, Kun Liu, Fei-Yu Shen, Yi-Chen Wu, Feng-Long Xie, Kun Xie, and Kai-Tuo Xu. 2024.
\newblock Fireredtts: A foundation text-to-speech framework for industry-level generative speech applications.
\newblock \emph{arXiv preprint}, abs/2409.03283.

\bibitem[{Guo et~al.(2021)Guo, Wen, Jiang, Luo, Zhang, Zhao, Li, Gong, Zou, Han et~al.}]{guo2021didispeech}
Tingwei Guo, Cheng Wen, Dongwei Jiang, Ne~Luo, Ruixiong Zhang, Shuaijiang Zhao, Wubo Li, Cheng Gong, Wei Zou, Kun Han, et~al. 2021.
\newblock Didispeech: A large scale mandarin speech corpus.
\newblock In \emph{ICASSP 2021-2021 IEEE International Conference on Acoustics, Speech and Signal Processing (ICASSP)}, pages 6968--6972. IEEE.

\bibitem[{He et~al.(2024)He, Shang, Wang, Li, Gu, Hua, Liu, Yang, Li, Shi, Wang, Chen, Zhang, and Wu}]{emilia}
Haorui He, Zengqiang Shang, Chaoren Wang, Xuyuan Li, Yicheng Gu, Hua Hua, Liwei Liu, Chen Yang, Jiaqi Li, Peiyang Shi, Yuancheng Wang, Kai Chen, Pengyuan Zhang, and Zhizheng Wu. 2024.
\newblock Emilia: An extensive, multilingual, and diverse speech dataset for large-scale speech generation.
\newblock In \emph{{SLT}}. {IEEE}.

\bibitem[{He et~al.(2025)He, Shang, Wang, Li, Gu, Hua, Liu, Yang, Li, Shi et~al.}]{emilia-large}
Haorui He, Zengqiang Shang, Chaoren Wang, Xuyuan Li, Yicheng Gu, Hua Hua, Liwei Liu, Chen Yang, Jiaqi Li, Peiyang Shi, et~al. 2025.
\newblock Emilia: A large-scale, extensive, multilingual, and diverse dataset for speech generation.
\newblock \emph{arXiv preprint}, 2501.15907.

\bibitem[{Hu et~al.(2024)Hu, Chen, Wang, Chng, and Zhang}]{rio}
Yuchen Hu, Chen Chen, Siyin Wang, Eng~Siong Chng, and Chao Zhang. 2024.
\newblock Robust zero-shot text-to-speech synthesis with reverse inference optimization.
\newblock \emph{arXiv preprint}, abs/2407.02243.

\bibitem[{Hussain et~al.(2025)Hussain, Neekhara, Yang, Casanova, Ghosh, Desta, Fejgin, Valle, and Li}]{koel-tts}
Shehzeen Hussain, Paarth Neekhara, Xuesong Yang, Edresson Casanova, Subhankar Ghosh, Mikyas~T. Desta, Roy Fejgin, Rafael Valle, and Jason Li. 2025.
\newblock Koel-tts: Enhancing llm based speech generation with preference alignment and classifier free guidance.
\newblock \emph{arXiv preprint}, abs/2502.05236.

\bibitem[{Ju et~al.(2024)Ju, Wang, Shen, Tan, Xin, Yang, Liu, Leng, Song, Tang et~al.}]{naturalspeech3}
Zeqian Ju, Yuancheng Wang, Kai Shen, Xu~Tan, Detai Xin, Dongchao Yang, Eric Liu, Yichong Leng, Kaitao Song, Siliang Tang, et~al. 2024.
\newblock Naturalspeech 3: Zero-shot speech synthesis with factorized codec and diffusion models.
\newblock In \emph{Forty-first International Conference on Machine Learning}.

\bibitem[{Kahn et~al.(2020)Kahn, Riviere, Zheng, Kharitonov, Xu, Mazar{\'e}, Karadayi, Liptchinsky, Collobert, Fuegen et~al.}]{kahn2020libri}
Jacob Kahn, Morgane Riviere, Weiyi Zheng, Evgeny Kharitonov, Qiantong Xu, Pierre-Emmanuel Mazar{\'e}, Julien Karadayi, Vitaliy Liptchinsky, Ronan Collobert, Christian Fuegen, et~al. 2020.
\newblock Libri-light: A benchmark for asr with limited or no supervision.
\newblock In \emph{ICASSP 2020-2020 IEEE International Conference on Acoustics, Speech and Signal Processing (ICASSP)}, pages 7669--7673. IEEE.

\bibitem[{Le et~al.(2023)Le, Vyas, Shi, Karrer, Sari, Moritz, Williamson, Manohar, Adi, Mahadeokar, and Hsu}]{voicebox}
Matthew Le, Apoorv Vyas, Bowen Shi, Brian Karrer, Leda Sari, Rashel Moritz, Mary Williamson, Vimal Manohar, Yossi Adi, Jay Mahadeokar, and Wei{-}Ning Hsu. 2023.
\newblock Voicebox: Text-guided multilingual universal speech generation at scale.
\newblock In \emph{NeurIPS}.

\bibitem[{Li et~al.(2025{\natexlab{a}})Li, Lin, Li, Huang, Wang, Wang, Zhan, and Wu}]{dualcodec}
Jiaqi Li, Xiaolong Lin, Zhekai Li, Shixi Huang, Yuancheng Wang, Chaoren Wang, Zhenpeng Zhan, and Zhizheng Wu. 2025{\natexlab{a}}.
\newblock Dualcodec: A low-frame-rate, semantically-enhanced neural audio codec for speech generation.
\newblock In \emph{{INTERSPEECH}}. {ISCA}.

\bibitem[{Li et~al.(2025{\natexlab{b}})Li, Zhang, Wang, He, Wang, Wang, Liao, Ao, Xie, Huang, Zhang, and Wu}]{amphion_v0.2}
Jiaqi Li, Xueyao Zhang, Yuancheng Wang, Haorui He, Chaoren Wang, Li~Wang, Huan Liao, Junyi Ao, Zeyu Xie, Yiqiao Huang, Junan Zhang, and Zhizheng Wu. 2025{\natexlab{b}}.
\newblock Overview of the amphion toolkit (v0.2).
\newblock \emph{arXiv preprint arXiv:2501.15442}.

\bibitem[{Li et~al.(2024{\natexlab{a}})Li, Xu, and Yu}]{li2024rl}
Ziniu Li, Tian Xu, and Yang Yu. 2024{\natexlab{a}}.
\newblock When is rl better than dpo in rlhf? a representation and optimization perspective.
\newblock In \emph{The Second Tiny Papers Track at ICLR 2024}.

\bibitem[{Li et~al.(2024{\natexlab{b}})Li, Xu, Zhang, Lin, Yu, Sun, and Luo}]{li2023rema4}
Ziniu Li, Tian Xu, Yushun Zhang, Zhihang Lin, Yang Yu, Ruoyu Sun, and Zhi-Quan Luo. 2024{\natexlab{b}}.
\newblock Remax: A simple, effective, and efficient reinforcement learning method for aligning large language models.
\newblock In \emph{Forty-first International Conference on Machine Learning}.

\bibitem[{Liao et~al.(2024)Liao, Han, Yang, Du, Yang, Xu, Xu, Liu, Lu, and Li}]{baton}
Huan Liao, Haonan Han, Kai Yang, Tianjiao Du, Rui Yang, Qinmei Xu, Zunnan Xu, Jingquan Liu, Jiasheng Lu, and Xiu Li. 2024.
\newblock {BATON:} aligning text-to-audio model using human preference feedback.
\newblock In \emph{{IJCAI}}, pages 4542--4550. ijcai.org.

\bibitem[{Lipman et~al.(2023)Lipman, Chen, Ben{-}Hamu, Nickel, and Le}]{flow-matching}
Yaron Lipman, Ricky T.~Q. Chen, Heli Ben{-}Hamu, Maximilian Nickel, and Matthew Le. 2023.
\newblock Flow matching for generative modeling.
\newblock In \emph{{ICLR}}. OpenReview.net.

\bibitem[{Majumder et~al.(2024)Majumder, Hung, Ghosal, Hsu, Mihalcea, and Poria}]{tango2}
Navonil Majumder, Chia{-}Yu Hung, Deepanway Ghosal, Wei{-}Ning Hsu, Rada Mihalcea, and Soujanya Poria. 2024.
\newblock Tango 2: Aligning diffusion-based text-to-audio generations through direct preference optimization.
\newblock In \emph{{ACM} Multimedia}, pages 564--572. {ACM}.

\bibitem[{Neekhara et~al.(2024)Neekhara, Hussain, Ghosh, Li, Valle, Badlani, and Ginsburg}]{robustness-problem}
Paarth Neekhara, Shehzeen Hussain, Subhankar Ghosh, Jason Li, Rafael Valle, Rohan Badlani, and Boris Ginsburg. 2024.
\newblock Improving robustness of llm-based speech synthesis by learning monotonic alignment.
\newblock In \emph{{INTERSPEECH}}. {ISCA}.

\bibitem[{Ouyang et~al.(2022)Ouyang, Wu, Jiang, Almeida, Wainwright, Mishkin, Zhang, Agarwal, Slama, Ray, Schulman, Hilton, Kelton, Miller, Simens, Askell, Welinder, Christiano, Leike, and Lowe}]{instructgpt}
Long Ouyang, Jeffrey Wu, Xu~Jiang, Diogo Almeida, Carroll~L. Wainwright, Pamela Mishkin, Chong Zhang, Sandhini Agarwal, Katarina Slama, Alex Ray, John Schulman, Jacob Hilton, Fraser Kelton, Luke Miller, Maddie Simens, Amanda Askell, Peter Welinder, Paul~F. Christiano, Jan Leike, and Ryan Lowe. 2022.
\newblock Training language models to follow instructions with human feedback.
\newblock In \emph{NeurIPS}.

\bibitem[{Panayotov et~al.(2015)Panayotov, Chen, Povey, and Khudanpur}]{librispeech}
Vassil Panayotov, Guoguo Chen, Daniel Povey, and Sanjeev Khudanpur. 2015.
\newblock Librispeech: an asr corpus based on public domain audio books.
\newblock In \emph{2015 IEEE international conference on acoustics, speech and signal processing (ICASSP)}, pages 5206--5210. IEEE.

\bibitem[{Peng et~al.(2024)Peng, Huang, Li, Mohamed, and Harwath}]{voicecraft}
Puyuan Peng, Po-Yao Huang, Shang-Wen Li, Abdelrahman Mohamed, and David Harwath. 2024.
\newblock Voicecraft: Zero-shot speech editing and text-to-speech in the wild.
\newblock \emph{arXiv preprint arXiv:2403.16973}.

\bibitem[{Radford et~al.(2023)Radford, Kim, Xu, Brockman, McLeavey, and Sutskever}]{whisper}
Alec Radford, Jong~Wook Kim, Tao Xu, Greg Brockman, Christine McLeavey, and Ilya Sutskever. 2023.
\newblock Robust speech recognition via large-scale weak supervision.
\newblock In \emph{International conference on machine learning}, pages 28492--28518. PMLR.

\bibitem[{Rafailov et~al.(2023)Rafailov, Sharma, Mitchell, Manning, Ermon, and Finn}]{dpo}
Rafael Rafailov, Archit Sharma, Eric Mitchell, Christopher~D. Manning, Stefano Ermon, and Chelsea Finn. 2023.
\newblock Direct preference optimization: Your language model is secretly a reward model.
\newblock In \emph{NeurIPS}.

\bibitem[{Saeki et~al.(2022)Saeki, Xin, Nakata, Koriyama, Takamichi, and Saruwatari}]{utmos}
Takaaki Saeki, Detai Xin, Wataru Nakata, Tomoki Koriyama, Shinnosuke Takamichi, and Hiroshi Saruwatari. 2022.
\newblock {UTMOS:} utokyo-sarulab system for voicemos challenge 2022.
\newblock In \emph{{INTERSPEECH}}, pages 4521--4525. {ISCA}.

\bibitem[{Sahoo et~al.(2024)Sahoo, Meharia, Ghosh, Saha, Jain, and Chadha}]{hallucination-survey}
Pranab Sahoo, Prabhash Meharia, Akash Ghosh, Sriparna Saha, Vinija Jain, and Aman Chadha. 2024.
\newblock A comprehensive survey of hallucination in large language, image, video and audio foundation models.
\newblock In \emph{{EMNLP} (Findings)}, pages 11709--11724. Association for Computational Linguistics.

\bibitem[{Shen et~al.(2024)Shen, Ju, Tan, Liu, Leng, He, Qin, Zhao, and Bian}]{naturalspeech2}
Kai Shen, Zeqian Ju, Xu~Tan, Eric Liu, Yichong Leng, Lei He, Tao Qin, Sheng Zhao, and Jiang Bian. 2024.
\newblock Naturalspeech 2: Latent diffusion models are natural and zero-shot speech and singing synthesizers.
\newblock In \emph{{ICLR}}. OpenReview.net.

\bibitem[{Skalse et~al.(2022)Skalse, Howe, Krasheninnikov, and Krueger}]{reward-hacking}
Joar Skalse, Nikolaus Howe, Dmitrii Krasheninnikov, and David Krueger. 2022.
\newblock Defining and characterizing reward gaming.
\newblock In \emph{NeurIPS}, volume~35, pages 9460--9471.

\bibitem[{Tan(2023)}]{tts-book-tanxu}
Xu~Tan. 2023.
\newblock \emph{Neural Text-to-Speech Synthesis}.
\newblock Springer.

\bibitem[{Tian et~al.(2024)Tian, Zhang, Shi, Zhang, Yu, Watanabe, and Yu}]{tianjinchuan-pa}
Jinchuan Tian, Chunlei Zhang, Jiatong Shi, Hao Zhang, Jianwei Yu, Shinji Watanabe, and Dong Yu. 2024.
\newblock Preference alignment improves language model-based {TTS}.
\newblock \emph{arXiv preprint}, abs/2409.12403.

\bibitem[{Wallace et~al.(2024)Wallace, Dang, Rafailov, Zhou, Lou, Purushwalkam, Ermon, Xiong, Joty, and Naik}]{wallace2024diffusion}
Bram Wallace, Meihua Dang, Rafael Rafailov, Linqi Zhou, Aaron Lou, Senthil Purushwalkam, Stefano Ermon, Caiming Xiong, Shafiq Joty, and Nikhil Naik. 2024.
\newblock Diffusion model alignment using direct preference optimization.
\newblock In \emph{Proceedings of the IEEE/CVF Conference on Computer Vision and Pattern Recognition}, pages 8228--8238.

\bibitem[{Wang et~al.(2023)Wang, Chen, Wu, Zhang, Zhou, Liu, Chen, Liu, Wang, Li, He, Zhao, and Wei}]{valle}
Chengyi Wang, Sanyuan Chen, Yu~Wu, Ziqiang Zhang, Long Zhou, Shujie Liu, Zhuo Chen, Yanqing Liu, Huaming Wang, Jinyu Li, Lei He, Sheng Zhao, and Furu Wei. 2023.
\newblock Neural codec language models are zero-shot text to speech synthesizers.
\newblock \emph{arXiv preprint}, abs/2301.02111.

\bibitem[{Wang et~al.(2025{\natexlab{a}})Wang, Zhan, Liu, Zeng, Guo, Zheng, Zhang, Zhang, Zhang, and Wu}]{maskgct}
Yuancheng Wang, Haoyue Zhan, Liwei Liu, Ruihong Zeng, Haotian Guo, Jiachen Zheng, Qiang Zhang, Xueyao Zhang, Shunsi Zhang, and Zhizheng Wu. 2025{\natexlab{a}}.
\newblock Maskgct: Zero-shot text-to-speech with masked generative codec transformer.
\newblock In \emph{{ICLR}}. OpenReview.net.

\bibitem[{Wang et~al.(2025{\natexlab{b}})Wang, Zheng, Zhang, Zhang, Liao, and Wu}]{wang2025metis}
Yuancheng Wang, Jiachen Zheng, Junan Zhang, Xueyao Zhang, Huan Liao, and Zhizheng Wu. 2025{\natexlab{b}}.
\newblock Metis: A foundation speech generation model with masked generative pre-training.
\newblock \emph{arXiv preprint arXiv:2502.03128}.

\bibitem[{Weng(2024)}]{reward-hacking-wenglilian}
Lilian Weng. 2024.
\newblock \href {https://lilianweng.github.io/posts/2024-11-28-reward-hacking/} {Reward hacking in reinforcement learning.}
\newblock \emph{lilianweng.github.io}.

\bibitem[{Xiong et~al.(2024)Xiong, Dong, Ye, Wang, Zhong, Ji, Jiang, and Zhang}]{iterative-dpo}
Wei Xiong, Hanze Dong, Chenlu Ye, Ziqi Wang, Han Zhong, Heng Ji, Nan Jiang, and Tong Zhang. 2024.
\newblock Iterative preference learning from human feedback: Bridging theory and practice for {RLHF} under kl-constraint.
\newblock In \emph{{ICML}}. OpenReview.net.

\bibitem[{Xu et~al.(2023)Xu, Liu, Wu, Tong, Li, Ding, Tang, and Dong}]{imagereward}
Jiazheng Xu, Xiao Liu, Yuchen Wu, Yuxuan Tong, Qinkai Li, Ming Ding, Jie Tang, and Yuxiao Dong. 2023.
\newblock Imagereward: Learning and evaluating human preferences for text-to-image generation.
\newblock In \emph{NeurIPS}.

\bibitem[{Yang et~al.(2024{\natexlab{a}})Yang, Yang, Hui, Zheng, Yu, Zhou, Li, Li, Liu, Huang, Dong, Wei, Lin, Tang, Wang, Yang, Tu, Zhang, Ma, Yang, Xu, Zhou, Bai, He, Lin, Dang, Lu, Chen, Yang, Li, Xue, Ni, Zhang, Wang, Peng, Men, Gao, Lin, Wang, Bai, Tan, Zhu, Li, Liu, Ge, Deng, Zhou, Ren, Zhang, Wei, Ren, Liu, Fan, Yao, Zhang, Wan, Chu, Liu, Cui, Zhang, Guo, and Fan}]{qwen2}
An~Yang, Baosong Yang, Binyuan Hui, Bo~Zheng, Bowen Yu, Chang Zhou, Chengpeng Li, Chengyuan Li, Dayiheng Liu, Fei Huang, Guanting Dong, Haoran Wei, Huan Lin, Jialong Tang, Jialin Wang, Jian Yang, Jianhong Tu, Jianwei Zhang, Jianxin Ma, Jianxin Yang, Jin Xu, Jingren Zhou, Jinze Bai, Jinzheng He, Junyang Lin, Kai Dang, Keming Lu, Keqin Chen, Kexin Yang, Mei Li, Mingfeng Xue, Na~Ni, Pei Zhang, Peng Wang, Ru~Peng, Rui Men, Ruize Gao, Runji Lin, Shijie Wang, Shuai Bai, Sinan Tan, Tianhang Zhu, Tianhao Li, Tianyu Liu, Wenbin Ge, Xiaodong Deng, Xiaohuan Zhou, Xingzhang Ren, Xinyu Zhang, Xipin Wei, Xuancheng Ren, Xuejing Liu, Yang Fan, Yang Yao, Yichang Zhang, Yu~Wan, Yunfei Chu, Yuqiong Liu, Zeyu Cui, Zhenru Zhang, Zhifang Guo, and Zhihao Fan. 2024{\natexlab{a}}.
\newblock Qwen2 technical report.
\newblock \emph{arXiv preprint}, abs/2407.10671.

\bibitem[{Yang et~al.(2024{\natexlab{b}})Yang, Yang, Zhang, Hui, Zheng, Yu, Li, Liu, Huang, Wei, Lin, Yang, Tu, Zhang, Yang, Yang, Zhou, Lin, Dang, Lu, Bao, Yang, Yu, Li, Xue, Zhang, Zhu, Men, Lin, Li, Xia, Ren, Ren, Fan, Su, Zhang, Wan, Liu, Cui, Zhang, and Qiu}]{qwen2.5}
An~Yang, Baosong Yang, Beichen Zhang, Binyuan Hui, Bo~Zheng, Bowen Yu, Chengyuan Li, Dayiheng Liu, Fei Huang, Haoran Wei, Huan Lin, Jian Yang, Jianhong Tu, Jianwei Zhang, Jianxin Yang, Jiaxi Yang, Jingren Zhou, Junyang Lin, Kai Dang, Keming Lu, Keqin Bao, Kexin Yang, Le~Yu, Mei Li, Mingfeng Xue, Pei Zhang, Qin Zhu, Rui Men, Runji Lin, Tianhao Li, Tingyu Xia, Xingzhang Ren, Xuancheng Ren, Yang Fan, Yang Su, Yichang Zhang, Yu~Wan, Yuqiong Liu, Zeyu Cui, Zhenru Zhang, and Zihan Qiu. 2024{\natexlab{b}}.
\newblock Qwen2.5 technical report.
\newblock \emph{arXiv preprint}, abs/2412.15115.

\bibitem[{Yao et~al.(2025)Yao, Yang, Pan, Feng, Ning, Ye, Zhou, and Xie}]{fpo}
Jixun Yao, Yuguang Yang, Yu~Pan, Yuan Feng, Ziqian Ning, Jianhao Ye, Hongbin Zhou, and Lei Xie. 2025.
\newblock Fine-grained preference optimization improves zero-shot text-to-speech.
\newblock \emph{arXiv preprint}, abs/2502.02950.

\bibitem[{Zeghidour et~al.(2021)Zeghidour, Luebs, Omran, Skoglund, and Tagliasacchi}]{soundstream}
Neil Zeghidour, Alejandro Luebs, Ahmed Omran, Jan Skoglund, and Marco Tagliasacchi. 2021.
\newblock Soundstream: An end-to-end neural audio codec.
\newblock \emph{IEEE/ACM Transactions on Audio, Speech, and Language Processing}, 30:495--507.

\bibitem[{Zhang et~al.(2024{\natexlab{a}})Zhang, Li, Li, Zhang, Wang, Zhou, and Qiu}]{speechalign}
Dong Zhang, Zhaowei Li, Shimin Li, Xin Zhang, Pengyu Wang, Yaqian Zhou, and Xipeng Qiu. 2024{\natexlab{a}}.
\newblock Speechalign: Aligning speech generation to human preferences.
\newblock In \emph{NeurIPS}.

\bibitem[{Zhang et~al.(2024{\natexlab{b}})Zhang, Lin, Bai, and Mei}]{zhang2024negative}
Ruiqi Zhang, Licong Lin, Yu~Bai, and Song Mei. 2024{\natexlab{b}}.
\newblock Negative preference optimization: From catastrophic collapse to effective unlearning.
\newblock \emph{arXiv preprint arXiv:2404.05868}.

\bibitem[{Zhang et~al.(2024{\natexlab{c}})Zhang, Xue, Gu, Wang, Li, He, Wang, Song, Chen, Fang, Chen, Zhang, Tang, Zou, Wang, Han, Chen, Li, and Wu}]{amphion}
Xueyao Zhang, Liumeng Xue, Yicheng Gu, Yuancheng Wang, Jiaqi Li, Haorui He, Chaoren Wang, Ting Song, Xi~Chen, Zihao Fang, Haopeng Chen, Junan Zhang, Tze~Ying Tang, Lexiao Zou, Mingxuan Wang, Jun Han, Kai Chen, Haizhou Li, and Zhizheng Wu. 2024{\natexlab{c}}.
\newblock Amphion: An open-source audio, music and speech generation toolkit.
\newblock In \emph{{IEEE} Spoken Language Technology Workshop, {SLT} 2024}.

\bibitem[{Zhang et~al.(2025)Zhang, Zhang, Peng, Tang, Manohar, Liu, Hwang, Li, Wang, Chan, Huang, Wu, and Ma}]{vevo}
Xueyao Zhang, Xiaohui Zhang, Kainan Peng, Zhenyu Tang, Vimal Manohar, Yingru Liu, Jeff Hwang, Dangna Li, Yuhao Wang, Julian Chan, Yuan Huang, Zhizheng Wu, and Mingbo Ma. 2025.
\newblock Vevo: Controllable zero-shot voice imitation with self-supervised disentanglement.
\newblock In \emph{{ICLR}}. OpenReview.net.

\bibitem[{Zhang et~al.(2024{\natexlab{d}})Zhang, Pan, Guo, Li, Zhu, Wang, Xu, Lu, Hong, Wang, Zhang, He, Jiang, Chen, Yang, Zhou, Cheng, and Zhao}]{gtsinger}
Yu~Zhang, Changhao Pan, Wenxiang Guo, Ruiqi Li, Zhiyuan Zhu, Jialei Wang, Wenhao Xu, Jingyu Lu, Zhiqing Hong, Chuxin Wang, Lichao Zhang, Jinzheng He, Ziyue Jiang, Yuxin Chen, Chen Yang, Jiecheng Zhou, Xinyu Cheng, and Zhou Zhao. 2024{\natexlab{d}}.
\newblock Gtsinger: {A} global multi-technique singing corpus with realistic music scores for all singing tasks.
\newblock In \emph{NeurIPS}.

\end{thebibliography}

\appendix

\section{Construction Details of INTP}

\subsection{Prompt Construction}
\label{sec:appendix-prompt-construction}

We construct English and Chinese prompt data, both based on the Emilia-Large dataset~\cite{emilia,emilia-large}, which contains diverse real-world speech data across various topics, recording scenarios, and speaking styles. 

\paragraph{Reference Speech}

We perform stratified sampling on Emilia-Large's speech data based on its metadata such as topics and tags to cover diverse acoustic conditions. Considering the memory constraints of existing zero-shot TTS models during inference, we only select samples with durations not exceeding 12 seconds.

\paragraph{Target Text}\label{sec:appendix-prompt-target-text-construction}
Similarly to reference speech, we perform stratified sampling based on Emilia-Large's metadata to cover diverse semantic topics. We select speech samples with durations between 5 and 22 seconds, and use their corresponding textual transcriptions as the target text data source.

We utilize DeepSeek V3~\cite{deepseek-r1} to preprocess the sampled textual transcriptions, such as typo correction and punctuation mark normalization, and use the processed text as regular text in INTP. Specifically, we use the following instruction for DeepSeek V3 to conduct text preprocessing:

\begin{promptbox}
\textit{\textbf{System Prompt:} }\\
I obtained a text from an audio file based on some ASR models. Please help me clean it up (e.g., correct typos, add proper punctuation marks, and make the sentences semantically coherent). Note: (1) You can modify, add, or replace words that better fit the context to ensure semantic coherence. (2) Please only return the cleaned-up result without any explanation.

\medskip \textit{\textbf{User Prompt (Example):} }\\
a panda eats shoes and leaves

\medskip \textit{\textbf{System Output (Example):} }\\
A panda eats shoots and leaves.
\end{promptbox}

Furthermore, we employ DeepSeek V3 to transform the regular text into different types. To generate Chinese-English-mixed code-switching texts:

\begin{promptbox}
\textit{\textbf{System Prompt:} }\\
\begin{CJK*}{UTF8}{gbsn}
    请你把这句话，转换成一个中文、英文混合的~code-switching~版本。注意：你只需要返回给我转换后的结果，不需要任何解释。
\end{CJK*}

\medskip \textit{\textbf{User Prompt (Example):} }\\
A panda eats shoots and leaves.

\medskip \textit{\textbf{System Output (Example):} }\\
\begin{CJK*}{UTF8}{gbsn}
    熊猫吃~shoots~和~leaves。
\end{CJK*}

\end{promptbox}

To generate punctuation-perturbed texts:

\begin{promptbox}
\textit{\textbf{System Prompt:} }\\
\begin{CJK*}{UTF8}{gbsn}
    假设你是一个~Text To Speech (TTS)~领域的专家，现在，让我们对一个~TTS~系统进行攻击。具体地：我输入一个文本，请你修改这条文本里面的若干词语，从而使~TTS~系统更容易出错。例如：你可以修改为把某些字修改为容易读错的形近字、把多音字做替换，等等，但你不要增加和删除原有的文本。注意：你只需要返回给我转换后的结果，不需要任何解释。\\
    
    例子1: \\
    【我的输入】我今天很高兴\\
    【你的输出】窝锦添狠搞醒\\

    例子2: \\
    【我的输入】目前，爱心人士正在种作寄养的小猫已经五个月大了。而本人的种作寄养申请单需要进一步审核。为了避免小猫多次转手，治疗者们对小猫的种作寄养提出了严格要求：申请人需年满二十三岁。\\
    【你的输出】幕前，爱信人士正在重作寄扬的削猫已经伍个月大了。而本人的重作寄扬神情但需要进一步审核。为了闭面削猫多次转售，治理者们对削猫的重作寄扬提出了阉割要求：申情人需年慢贰拾叁岁。\\

    例子3: \\
    【我的输入】And the idea of standing all by himself in a crowded market, to be pushed and hired by some big, strange farmer, was very disagreeable. Why not sing that high note and grow potatoes?\\
    【你的输出】And the eye dear of standing awl bye himself in a crowd dead market, two bee pushed and high red buy sum big, strange far mer, was vary dis agreeable. Y knot sing that hi note and grow poe eight toes?\\
\end{CJK*}

\medskip \textit{\textbf{User Prompt (Example):} }\\
A panda eats shoots and leaves.

\medskip \textit{\textbf{System Output (Example):} }\\
A pan duh eights shots n leafs.

\end{promptbox}

To generate repeated text and punctuation-perturbed text, we leverage DeepSeek V3 to create executable Python scripts that implement rule-based word repetition and random punctuation modification. These scripts will be included in our future open-source repository.

\paragraph{Combination between Speech and Text}

Based on the language of reference speech and target text data, we design four balanced combination categories: monolingual combinations (\textit{en2en} and \textit{zh2zh}) and cross-lingual combinations (\textit{zh2en} and \textit{en2zh}), where \textit{zh2en} denotes Chinese reference speech with English target text, and similarly for others. For each text type shown in Table~\ref{tab:intp-stats} (Regular, Repeated, Code-Switching, Pronunciation-perturbed, and Punctuation-perturbed), we construct 12K prompts.

\subsection{Model Selection}\label{sec:appendix-model-selection}

\begin{itemize}[itemsep=0ex,leftmargin=2ex]
    \item \textbf{ARS}~\cite{maskgct}: We use the original checkpoint (pre-trained on Emilia) provided by the authors.
    \item \textbf{F5-TTS}~\cite{f5tts}: We use the officially released checkpoint\footnote{\href{https://huggingface.co/SWivid/F5-TTS}{https://huggingface.co/SWivid/F5-TTS}} for INTP data generation.
    \item \textbf{MaskGCT}~\cite{maskgct}: We use the officially released checkpoint\footnote{\href{https://huggingface.co/amphion/MaskGCT}{https://huggingface.co/amphion/MaskGCT}}~\cite{amphion,amphion_v0.2} for INTP data generation.
\end{itemize}

In addition to these three models used for INTP construction, we also investigate INTP's effectiveness on \textbf{CosyVoice 2} and \textbf{Ints}. For CosyVoice 2, we conduct alignment experiments using its officially released checkpoint\footnote{\href{https://github.com/FunAudioLLM/CosyVoice}{https://github.com/FunAudioLLM/CosyVoice}} as the base model. Details of the pre-trained models of Ints are provided in Appendix~\ref{sec:appendix-detail-ins}.

\subsection{Preference Pairs Construction}\label{sec:appendix-preference-pairs}

\subsubsection{Intra Pair}\label{sec:appendix-intp-intra-pair}

For each model and prompt, we perform five samplings and construct intra pairs based on their WER comparisons. To maximize the performance gap between positive and negative samples, we employ two strategies. First, we use diverse hyperparameters during the five generations to increase sample diversity, selecting the generation with the lowest WER as positive samples and the highest WER as negative samples. Second, we apply a threshold to filter out pairs where the WER gap between positive and negative samples is less than 6.0.

Specifically, for ARS's five samplings, we set top k to 20 and top p to 1.0, while using different temperatures of 0.4, 0.6, 0.8, 1.0, and 1.2. For F5-TTS and MaskGCT, we use the generated speech target duration as the sampling hyperparameter. Denoting the ``ground truth'' duration\footnote{Since we use Emilia-Large's transcription data as target text in our prompt construction process (Appendix~\ref{sec:appendix-prompt-target-text-construction}), we refer to the original speech duration corresponding to this transcription as the ``ground truth'' duration.} as $d$, we employ five different duration parameters: 0.8$d$, 0.9$d$, 1.0$d$, 1.1$d$, and 1.2$d$.

\subsubsection{Inter Pair}\label{sec:appendix-intp-inter-pair}

We construct inter pairs based on the intra pairs established in Appendix~\ref{sec:appendix-intp-intra-pair}. For a given prompt, we denote model A's intra pair as $(y^w_A, y^l_A)$ and model B's intra pair as $(y^w_B, y^l_B)$. We construct inter pairs through three types of comparisons: between $y^w_A$ and $y^w_B$, between $y^w_A$ and $y^l_B$, and between $y^l_A$ and $y^w_B$. Note that we exclude comparisons between $y^l_A$ and $y^l_B$ to ensure high quality of positive samples. We apply the same WER threshold as in Appendix~\ref{sec:appendix-intp-intra-pair} to filter out pairs with small performance gaps.

\subsubsection{Perturbed Pair}\label{sec:appendix-intp-perturbed-pair}

The instructions used to prompt DeepSeek V3 for obtaining pronunciation-perturbed and punctuation-perturbed texts are shown in Appendix~\ref{sec:appendix-prompt-target-text-construction}. Specifically, we only use data from INTP's regular text domain to construct perturbed pairs.

\subsection{Human Verification}\label{sec:appendix-intp-human-verficiation}

\begin{table}[t]
\begin{center}
\resizebox{0.8\columnwidth}{!}{%
\begin{threeparttable}
    \begin{tabular}{ccccc}
        \toprule
         \makecell[c]{\textbf{A+2}} & \makecell[c]{\textbf{A+1}} & \makecell[c]{\textbf{Tie}} & \makecell[c]{\textbf{B+1}} & \makecell[c]{\textbf{B+2}} \\ \midrule
         10.9\% & 29.0\% & 15.0\% & 32.4\% & 12.6\% \\
         \bottomrule
    \end{tabular}
    \begin{tablenotes}
        \footnotesize{
            \item[*] For each pair, we present the two samples to human raters in random order, labeled as A and B. A+2 indicates that sample A's naturalness is significantly better than B, while A+1 indicates that sample A is moderately better than B, similar for B+2 and B+1. Tie indicates no perceptible difference.
        }
    \end{tablenotes}
\end{threeparttable}
}
\caption{Human naturalness preference for 1,000 pairs from INTP regular text domain.}
\label{tab:human-naturalness-verify}
\end{center}
\end{table}

\begin{table}[t]
\begin{center}
\resizebox{\columnwidth}{!}{%
\begin{threeparttable}
    \begin{tabular}{c|ccc}
        \toprule
         & \textbf{\makecell[c]{Naturalness\\Winner}} & \textbf{\makecell[c]{Naturalness\\Tie}} & \textbf{\makecell[c]{Naturalness\\Loser}} \\ \midrule
         \textbf{INTP winner} & 72\% & 15\% & 13\% \\
         \bottomrule
    \end{tabular}
\end{threeparttable}
}
\caption{Agreement between INTP preference and human naturalness preference.}
\label{tab:human-naturalness-compare-wer}
\end{center}
\end{table}

In Section~\ref{sec:intp-human-verify}, we evaluated INTP's alignment with human intelligibility perception. In this section, we investigate the alignment between INTP and human naturalness preferences. Specifically, we design a \textbf{naturalness preference} annotation task (Appendix~\ref{sec:appendix-sub-eval}). We randomly sample 1,000 pairs from INTP's regular text domain for human annotation, with results shown in Table~\ref{tab:human-naturalness-verify} and~\ref{tab:human-naturalness-compare-wer}. The results reveal two key findings: First, 85\% of INTP pairs exhibit distinguishable naturalness preferences (Tie rate of 15\% in Table~\ref{tab:human-naturalness-verify}). Additionally, INTP's preference determination shows strong agreement with human naturalness preferences (72\% agreement rate between INTP winners and naturalness winners in Table~\ref{tab:human-naturalness-compare-wer}). These results suggest that INTP can also serve as a foundation dataset for naturalness preference alignment in future research.

\section{Details of the Derivation}\label{sec:appendix-derivation}

\subsection{DPO for AR Models}\label{sec:appendix-derivation-ar}

Starting from Equation~\ref{eq:rl_obj}, \citet{dpo} demonstrate that the optimization problem admits a closed-form solution. Specifically, the optimal policy \( p_\theta^*(y | x) \) that maximizes the RL objective is given by:
\begin{equation}
    {\small
    \begin{aligned}
p_\theta^*(y | x) = \frac{1}{Z(x)} p_{\text{ref}}(y | x) \exp\left(\frac{1}{\beta} r(x, y)\right),
    \end{aligned}
    }
\label{eq:optimal_policy}
\end{equation}
where \( Z(x) \) is the partition function ensuring normalization. This establishes a direct relationship between the reward function and the policy:
\begin{equation}
    {\small
    \begin{aligned}
    r(x, y) = \beta \log \frac{p_\theta^*(y | x)}{p_{\text{ref}}(y | x)} + \beta \log Z(x).
    \end{aligned}
    }
\label{eq:reward_expression}
\end{equation}
Substituting this reward expression (Equation~\ref{eq:reward_expression}) into the reward modeling loss function (Equation~\ref{eq:reward_model}) leads the DPO loss (Equation~\ref{eq:dpo_ar}), which we represent here as:
\[
    {\small
    \begin{aligned}
\mathcal{L}_\text{DPO} = -\mathbb{E}_{\mathcal{D}} \left[ \log \sigma\left( \beta \left( \log \tfrac{p_\theta(y_w | x)}{p_{\text{ref}}(y_w | x)} - \log \tfrac{p_\theta(y_l | x)}{p_{\text{ref}}(y_l | x)} \right) \right) \right].
    \end{aligned}
    }
\]

\subsection{DPO for Flow-Matching Models}\label{sec:appendix-derivation-fm}

Starting from Equation~\ref{eq:rl_obj_fm}, which we represent here as:
\[{\small
\begin{aligned}
\max_{p_\theta} & \mathbb{E}_{y_1 \sim p_\theta(y_1|x), t, x}[r(y_1,x)] \\
& - \beta \mathbb{D}_{\text{KL}}[p_\theta(y_1|y_t,t,x) \| p_{\text{ref}}(y_1|y_t,t,x)].
\end{aligned}
}\]
Similar to the derivation in DPO~\cite{dpo} and~\citet{wallace2024diffusion}, we obtain the closed-form solution for the optimal policy as:
\begin{equation}
    {\scriptsize
    \begin{aligned}
    p_\theta^*(y_1 | y_t, t, x) = \frac{1}{Z(y_t, t, x)} p_{\text{ref}}(y_1 | y_t, t, x) \exp\left(\frac{1}{\beta} r(y_1, x)\right),
    \end{aligned}
    }
\end{equation}
where \( Z(y_t, t, x) \) is the partition function ensuring normalization. We can then express the reward model \( r(y_1, x) \) as:
\begin{equation}
    {\small
    \begin{aligned}
    r(y_1, x) = \beta \log \frac{p_\theta^*(y_1 | y_t, t, x)}{p_{\text{ref}}(y_1 | y_t, t, x)} + \beta \log Z(y_t, t, x).
    \end{aligned}
    }
    \label{eq:reward_expression_fm}
\end{equation}
Similarly, substituting this reward expression (Equation~\ref{eq:reward_expression_fm}) into the reward modeling loss function (Equation~\ref{eq:reward_model}) leads to the DPO loss for OT-FM (Equation~\ref{eq:dpo-fm}), which we represent here:
\[
{\scriptsize
\begin{aligned}
& \mathcal{L}_{\text{DPO-FM}} = -
\mathbb{E}_{(y_1^w, y_1^l, x) \sim \mathcal{D}, t} \\
& \log \sigma \left( \beta \left( \log \frac{p_\theta(y_1^w|y_t^w,t,x)}{p_{\text{ref}}(y_1^w|y_t^w,t,x)}
- \log \frac{p_\theta(y_1^l|y_t^l,t,x)}{p_{\text{ref}}(y_1^l|y_t^l,t,x)} \right) \right).
\end{aligned}
}
\]
Reviewing the training objective of OT-FM (Equation~\ref{eq:obj_fm}), we find that it is equivalent to fitting a Gaussian likelihood. In other words, the induced likelihood can be interpreted as:
\[
{\small
\begin{aligned}
p_\theta(y_1 \mid y_t, t, x) \propto \exp\!\left(-\frac{1}{\beta}\left\|v_\theta(y_t,t,x) - (y_1-y_0)\right\|_2^2\right),
\end{aligned}
}
\]
similarly, for the reference policy, we have:
\[
{\small
\begin{aligned}
p_{\mathrm{ref}}(y_1 \mid y_t, t, x) \propto \exp\!\left(-\frac{1}{\beta}\left\|v_{\mathrm{ref}}(y_t,t,x) - (y_1-y_0)\right\|_2^2\right).
\end{aligned}
}
\]
Here, \(\beta\) serves as an inverse temperature (or noise variance), and the normalization constants cancel out when taking the ratio.
By taking the logarithm of the ratio between the learned policy and the reference policy, we obtain:
\[
{\small
\begin{aligned}
\log\frac{p_\theta(y_1\mid y_t,t,x)}{p_{\mathrm{ref}}(y_1\mid y_t,t,x)} = & -\frac{1}{\beta} \Bigl( \|v_\theta(y_t,t,x) - (y_1-y_0)\|_2^2 \\
& - \|v_{\mathrm{ref}}(y_t,t,x) - (y_1-y_0)\|_2^2\Bigr).
\end{aligned}
}
\]
Multiplying both sides by \(\beta\) results in:
\[
{\small
\begin{aligned}
\beta\log\frac{p_\theta(y_1\mid y_t,t,x)}{p_{\mathrm{ref}}(y_1\mid y_t,t,x)} = & -\Bigl( \|v_\theta(y_t,t,x) - (y_1-y_0)\|_2^2 \\
& - \|v_{\mathrm{ref}}(y_t,t,x) - (y_1-y_0)\|_2^2 \Bigr).
\end{aligned}
}
\]
By substituting the log-ratio formulation into Equation~\ref{eq:dpo-fm}, we can transform the DPO loss for OT-FM into a form related to the velocity, as shown in Equation~\ref{eq:dpo-fm-velocity}, which is represented as:
\[
{\tiny
\begin{aligned}
& \mathcal{L}_{\text{DPO-FM}} = -\mathbb{E}_{(y_1^w, y_1^l, x) \sim \mathcal{D}, t} \log \sigma \Big( -\beta \\
& \Big( \left\| v_\theta(y^w_t, t, x) - (y_1^w - y_0^w) \right\|_2^2
- \left\| v_\text{ref}(y^w_t, t, x) - (y_1^w - y_0^w) \right\|_2^2 \Big) \\
& - \Big( \left\| v_\theta(y^l_t, t, x) - (y_1^l - y_0^l) \right\|_2^2
- \left\| v_\text{ref}(y^l_t, t, x) - (y_1^l -y_0^l) \right\|_2^2 \Big) \Big).
\end{aligned}
}
\]

\subsection{DPO for Masked Generative Models}\label{sec:appendix-derivation-mgm}


Similar to flow-matching, let \(p_\theta(y_0 \mid y_t, x)\) denote the policy to be optimized, and \(p_{\text{ref}}(y_0 \mid y_t, x)\) the reference policy. We can rewrite the RL objective for MGM as follows:
\begin{equation}
{\small
\begin{aligned}
\max_{p_\theta} \quad & \mathbb{E}_{y_0 \sim p_\theta(y_0|x), t, x} \left[ r(y_0, x) \right] \\
& - \beta \mathbb{D}_{\text{KL}} \left[ p_\theta(y_0|y_t,x) \,\|\, p_{\text{ref}}(y_0|y_t,x) \right].
\end{aligned}
}
\end{equation}
We can also derive the closed-form solution for the optimal policy:
\begin{equation}
{\small
\begin{aligned}
p_\theta^*(y_0 | y_t, x) = \frac{1}{Z(y_t, x)} p_{\text{ref}}(y_0 | y_t, x) \exp\left(\frac{1}{\beta} r(y_0, x)\right),
\end{aligned}
}
\end{equation}
and express the reward model as follows:
\begin{equation}
{\small
\begin{aligned}
r(y_0, x) = \beta \log \frac{p_\theta^*(y_0 | y_t, x)}{p_{\text{ref}}(y_0 | y_t, x)} + \beta \log Z(y_t, x),
\end{aligned}
}
\label{eq:reward_expression_mgm}
\end{equation}
where \( Z(y_t, x) \) is the partition function ensuring normalization.
Also, substituting this reward expression (Equation~\ref{eq:reward_expression_mgm}) into the reward modeling loss function (Equation~\ref{eq:reward_model}) leads to the DPO loss for MGM:
\begin{equation}
{\scriptsize
\begin{aligned}
& \mathcal{L}_{\text{DPO-MGM}} = -\mathbb{E}_{(y^w, y^l, x) \sim \mathcal{D}, t}  \\
& \log \sigma \left( \beta \left( \log \frac{p_\theta(y^w_0|y^w_t,x)}{p_{\text{ref}}(y^w_0|y^w_t,x)}
 - \log \frac{p_\theta(y^l_0|y^l_t,x)}{p_{\text{ref}}(y^l_0|y^l_t,x)} \right) \right).
\end{aligned}
}
\end{equation}
Here, \(y^w_t\) and \(y^l_t\) are masked versions of \(y^w_0\) and \(y^l_0\) generated via the mask schedule \(\gamma(t)\). Note that $p_\theta(y_0|y_t,x)$ corresponds to the sum of the log-probabilities of the unmasked tokens in the context of MGM.

\section{Ints: Intelligibility-enhanced Speech Language Model}\label{sec:appendix-detail-ins}

Ints is an \underline{int}elligibility-enhanced \underline{s}peech language model. 
It follows a two-stage generation paradigm like~\cite{seedtts, cosyvoice, maskgct}: in the first stage, it uses an AR model to generate discrete speech tokens, while in the second stage, it employs a flow matching model to generate mel-spectrograms from speech tokens. We use the first-layer tokens from DualCodec~\cite{dualcodec} as the modeling target for the first stage of Ints, due to its efficient compression representation (12.5Hz tokens for 24kHz speech) and rich semantic information. Particularly, the first-stage AR model is directly \textbf{initialized from a large language model} while extending the vocabulary to include speech tokens. The codebook size of speech tokens is 16,384. Specifically, in this work, we use the 3.8B \texttt{Phi-3.5-mini-instruct}\footnote{\href{https://huggingface.co/microsoft/Phi-3.5-mini-instruct}{https://huggingface.co/microsoft/Phi-3.5-mini-instruct}}~\cite{phi3}, motivated by scaling up model size and leveraging the rich textual semantic knowledge.

\subsection{TTS Instruction Design}

We format the input as a text-to-speech instruction concatenated with speech tokens. The input sequence is represented as: 
\[
\left[\boldsymbol{I}, \boldsymbol{T}, <|\text{startofspeech}|>, \boldsymbol{S}, <|\text{endofspeech}|> \right]
\]
where \(\boldsymbol{I}\) is the instruction prefix (e.g., ``Please speak the following text out loud''), and \(\boldsymbol{T}\) and \(\boldsymbol{S}\) denote the text and speech token sequences, respectively. The special tokens \(<|\text{startofspeech}|>\) and \(<|\text{endofspeech}|>\) mark the boundaries of the speech token sequence.

During the inference stage for zero-shot TTS, the input sequence is represented as:
\[
\left[\boldsymbol{I}, \boldsymbol{T}_{\text{prompt}}, \boldsymbol{T}_{\text{target}}, <|\text{startofspeech}|>, \boldsymbol{S}_{\text{prompt}} \right]
\]
to generate the target speech tokens \(\boldsymbol{S}_{\text{target}}\). Here, \(\boldsymbol{T}_{\text{prompt}}\), \(\boldsymbol{T}_{\text{target}}\), \(\boldsymbol{S}_{\text{prompt}}\) are placeholders for the prompt text, target text, and prompt speech tokens, respectively.

\subsection{Training data}\label{sec:appendix-ints-training-data}

We pre-train Ints on Emilia~\cite{emilia}, which consists of about 100K hours of multilingual data. Following this, we use INTP alignment to obtain Ints v1. Ints v1 is then used to generate new preference data, which are employed to train Ints v2 for iterative alignment. We select prompts from the repeated and code-switching samples of INTP, which can be considered a more challenging subset of prompts. For each prompt, we use the same INTP intra-pair pipeline in Appendix~\ref{sec:appendix-intp-intra-pair} to construct preference pairs.

\begin{table*}[t]
\begin{center}
\resizebox{\textwidth}{!}{
\begin{threeparttable}
    \begin{tabular}{c||rrc|rrc|rrc|rrc||rrc}
    \toprule
    \rowcolor{gray!30} \multicolumn{16}{c}{\textbf{\textit{On English Evaluation Samples}}} \\
    \midrule 
    \multirow{2}{*}{\textbf{Model}} & 
    \multicolumn{3}{c|}{\makecell[c]{\textbf{Regular (\textit{en})}}} &
    \multicolumn{3}{c|}{\textbf{Articulatory (\textit{en})}} &
    \multicolumn{3}{c|}{\textbf{Code-switching (\textit{en2mixed})}} &
    \multicolumn{3}{c||}{\textbf{Cross-lingual (\textit{zh2en})}} &
    \multicolumn{3}{c}{\textbf{Avg}} \\
    \cmidrule(lr){2-13} \cmidrule(lr){14-16}
     & \textbf{WER} & \textbf{SIM} & \textbf{UTMOS} & \textbf{WER} & \textbf{SIM} & \textbf{UTMOS} & \textbf{WER} & \textbf{SIM} & \textbf{UTMOS} & \textbf{WER} & \textbf{SIM} & \textbf{UTMOS} & \textbf{WER} & \textbf{SIM} & \textbf{UTMOS} \\
    \midrule 
    \multicolumn{1}{l||}{\textbf{ARS}~\cite{maskgct}} & 3.55 & 0.682 & 3.560 & 15.98 & 0.675 & 3.400 & 48.59 & 0.629 & 3.190 & 15.22 & 0.697 & 3.150 & 20.84 & 0.671 & 3.325 \\
    \cmidrule(lr){1-16}
    \multicolumn{1}{r||}{\textbf{w/ en2en}} & \textbf{1.96} & \textbf{0.697} & 3.690 & 13.42 & 0.685 & 3.570 & 35.18 & 0.641 & 3.270 & 8.19 & 0.692 & 3.300 & 14.19 & 0.679 & 3.458 \\
    \multicolumn{1}{r||}{\textbf{w/ zh2zh}} & 2.76 & 0.692 & 3.660 & 13.90 & \textbf{0.687} & 3.550 & 36.65 & 0.644 & 3.260 & 8.92 & 0.694 & 3.320 & 15.06 & 0.679 & 3.448 \\
    \multicolumn{1}{r||}{\textbf{w/ en2zh, zh2en}} & 2.32 & 0.694 & \textbf{3.700} & \textbf{11.78} & 0.684 & \textbf{3.580} & 35.17 & \textbf{0.645} & \textbf{3.290} & \textbf{7.00} & 0.700 & \textbf{3.330} & \textbf{14.07} & 0.681 & \textbf{3.475} \\
    \multicolumn{1}{r||}{\textbf{w/ all}} & 2.35 & 0.695 & 3.680 & 13.76 & 0.686 & 3.560 & \textbf{33.53} & 0.642 & 3.240 & 7.38 & \textbf{0.704} & 3.310 & 14.26 & \textbf{0.682} & 3.448 \\
    \midrule
    \rowcolor{gray!30} \multicolumn{16}{c}{\textbf{\textit{On Chinese Evaluation Samples}}} \\
    \midrule 
    \multirow{2}{*}{\textbf{Model}} & 
    \multicolumn{3}{c|}{\textbf{Regular (\textit{zh})}} &
    \multicolumn{3}{c|}{\textbf{Articulatory (\textit{zh})}} &
    \multicolumn{3}{c|}{\textbf{Code-switching (\textit{zh2mixed})}} &
    \multicolumn{3}{c||}{\textbf{Cross-lingual (\textit{en2zh})}} &
    \multicolumn{3}{c}{\textbf{Avg}} \\
    \cmidrule(lr){2-13} \cmidrule(lr){14-16}
     & \textbf{WER} & \textbf{SIM} & \textbf{UTMOS} & \textbf{WER} & \textbf{SIM} & \textbf{UTMOS} & \textbf{WER} & \textbf{SIM} & \textbf{UTMOS} & \textbf{WER} & \textbf{SIM} & \textbf{UTMOS} & \textbf{WER} & \textbf{SIM} & \textbf{UTMOS} \\
    \midrule 
    \multicolumn{1}{l||}{\textbf{ARS}~\cite{maskgct}} & 4.37 & 0.752 & 2.730 & 24.07 & 0.711 & 2.430 & 59.71 & 0.756 & 2.900 & 24.30 & 0.563 & 3.090 & 28.61 & 0.696 & 2.788 \\
    \cmidrule(lr){1-16}
    \multicolumn{1}{r||}{\textbf{w/ en2en}} & 2.68 & 0.761 & \textbf{2.760} & 21.68 & 0.727 & \textbf{2.530} & 48.84 & 0.757 & 2.990 & 12.48 & 0.566 & 3.140 & 21.42 & 0.703 & \textbf{2.855} \\
    \multicolumn{1}{r||}{\textbf{w/ zh2zh}} & \textbf{2.41} & 0.760 & 2.740 & \textbf{19.51} & \textbf{0.727} & 2.490 & 47.99 & 0.755 & \textbf{3.010} & 12.73 & 0.565 & 3.110 & 20.16 & 0.702 & 2.838 \\
    \multicolumn{1}{r||}{\textbf{w/ en2zh, zh2en}} & 2.49 & \textbf{0.762} & 2.740 & 22.92 & 0.715 & 2.490 & \textbf{41.00} & 0.757 & 3.000 & \textbf{11.76} & \textbf{0.573} & \textbf{3.160} & \textbf{19.54} & 0.702 & 2.848 \\
    \multicolumn{1}{r||}{\textbf{w/ all}} & 2.62 & 0.759 & 2.720 & 21.06 & 0.725 & 2.440 & 41.50 & \textbf{0.760} & 2.980 & 11.95 & 0.572 & 3.090 & 19.78 & \textbf{0.704} & 2.808 \\
    \bottomrule
    \end{tabular}
\end{threeparttable}
}
\caption{Effect of different languages within INTP for ARS. In these experiments, we use only the \textbf{Regular} part of INTP for training.}
\label{tab:expt-effect-lang}
\end{center}
\end{table*}

\section{Training Details}\label{sec:appendix-train-detail}

All of our experiments are conducted on 8 NVIDIA H100 80GB-GPUs. Unless stated otherwise, we use the AdamW optimizer with $\beta_1 = 0.9, \beta_2 = 0.999$ and train for one epoch. For each model, we provide more detailed information about the experiments:

\begin{itemize}[itemsep=0ex,leftmargin=2ex]
    \item \textbf{ARS}: We use a learning rate of $5e-6$  with a warmup of $4,000$ steps and an inverse square root learning scheduler. For DPO, we use the hyperparameter $\beta = 0.1$.
    \item \textbf{F5-TTS}: We use a learning rate of $8e-6$ with a warmup of $4,000$ steps and an inverse square root learning scheduler. For DPO, we use the hyperparameter $\beta = 1,000$.
    \item \textbf{MaskGCT}: We use a learning rate of $5e-6$  with a warmup of $4,000$ steps and an inverse square root learning scheduler. For DPO, we use the hyperparameter $\beta = 10$.
    \item \textbf{CosyVoice 2}: We use a learning rate of $5e-6$  with a warmup of $4,000$ steps and an inverse square root learning scheduler. For DPO, we use the hyperparameter $\beta = 0.1$.
    \item \textbf{Ints}: We use a learning rate of $5e-6$ with a warmup of $4,000$ steps and an inverse square root learning scheduler. For DPO, we use the hyperparameter $\beta = 0.1$. We use flash attention~\cite{flashattention} and bfloat16 for training. 
\end{itemize}

\section{Additional Experimental Results}\label{sec:appendix-more-results}

\subsection{Effect of Data across Different Languages within INTP}


We present the effect of different languages within INTP in Table~\ref{tab:expt-effect-lang}. The results reveal three key findings: (1) Data from all languages can contribute to improvements across diverse domains for ARS. (2) Interestingly, using only English post-training data (w/ en2en) could also improve performance on Chinese evaluation samples, and vice versa, demonstrating that the proposed alignment algorithm enhances the model's foundation capability in intelligibility. (3) Furthermore, we again observe the effectiveness of preference alignment's customized feature: when aiming to improve performance on cross-lingual cases, directly constructing data from the cross-lingual distribution yields the most significant gains.

\subsection{Effect of INTP Alignment for Unseen Languages}

\begin{table*}[t]
\begin{center}
\begin{threeparttable}
    \resizebox{.65\textwidth}{!}{
    \begin{tabular}{l|rc|rc|rc|rc}
    \toprule
    \multicolumn{1}{c|}{\multirow{2}{*}{\textbf{Model}}} & 
    \multicolumn{2}{c|}{\textbf{Japanese}} &
    \multicolumn{2}{c|}{\textbf{Korean}} & 
    \multicolumn{2}{c|}{\textbf{German}} &
    \multicolumn{2}{c}{\textbf{French}} \\
    \cmidrule(lr){2-9}
     & \textbf{WER} & \textbf{SIM} & \textbf{WER} & \textbf{SIM} & \textbf{WER} & \textbf{SIM} & \textbf{WER} & \textbf{SIM} \\
    \midrule
    \textbf{Ints}
        & 26.34 & 0.714
        & 31.67 & 0.708
        & 28.25 & 0.674
        & 54.53 & 0.545 \\
    \multicolumn{1}{r|}{\quad \textbf{w/ INTP}}
        & 21.82 & 0.718
        & 19.57 & 0.741
        & 21.20 & 0.676
        & 42.12 & 0.558 \\
    \bottomrule
    \end{tabular}
    }
\end{threeparttable}
\caption{Effect of INTP alignment for unseen languages.}
\label{tab:unseen_language}
\end{center}
\end{table*}

We conducted the additional evaluations on four unseen languages not covered by INTP. Specifically, we tested the Ints models before and after INTP alignment using Japanese, Korean, German, and French speech data from GTSinger~\cite{gtsinger} (a dataset not used in either pre-training or post-training). We constructed evaluation sets consisting of 500 samples for each language. The results in Table~\ref{tab:unseen_language} demonstrate that despite INTP containing only Chinese and English data, improvements in both WER and SIM metrics are observed across all four languages. We hypothesize that this generalization stems from our proposed intelligibility preference alignment method enhancing the model's fundamental capabilities in intelligibility such as the basic articulation and pronunciation.

\section{Evaluation Details}\label{sec:appendix-eval}

\subsection{Evaluation Data}\label{sec:appendix-eval-data}

Our evaluation sets are based on SeedTTS test-en and SeedTTS test-zh datasets\footnote{\href{https://github.com/BytedanceSpeech/seed-tts-eval}{https://github.com/BytedanceSpeech/seed-tts-eval}}. The SeedTTS test-en set includes 1,000 samples from the Common Voice dataset~\cite{ardila2019common}, while the SeedTTS test-zh set comprises 2,000 samples from the DiDiSpeech dataset~\cite{guo2021didispeech}.
We also provide the detailed distribution of our proposed sets in Table~\ref{tab:eval-data}.

\begin{table}[h]
\begin{center}
\resizebox{.9\columnwidth}{!}{%
    \begin{threeparttable}
        \begin{tabular}{c|c|c|c|c|c}
            \toprule
             & \multicolumn{4}{c|}{\textbf{Languages}} & \textbf{\#Total} \\
            \midrule
            \multirow{2}{*}{\textbf{\makecell[c]{Regular}}} & \multicolumn{2}{c|}{\textbf{en}} & \multicolumn{2}{c|}{\textbf{zh}} & \multirow{2}{*}{3,000} \\ \cmidrule(lr){2-5}
             & \multicolumn{2}{c|}{1,000} & \multicolumn{2}{c|}{2,000} & \\
            \midrule
            \multirow{2}{*}{\textbf{\makecell[c]{Articulatory}}} & \multicolumn{2}{c|}{\textbf{en}} & \multicolumn{2}{c|}{\textbf{zh}} & \multirow{2}{*}{800} \\ \cmidrule(lr){2-5}
             & \multicolumn{2}{c|}{400} & \multicolumn{2}{c|}{400} & \\
            \midrule
            \multirow{2}{*}{\textbf{\makecell[c]{Code-switching}}} & \multicolumn{2}{c|}{\textbf{en2mixed}} & \multicolumn{2}{c|}{\textbf{zh2mixed}} & \multirow{2}{*}{1,000} \\ \cmidrule(lr){2-5}
             & \multicolumn{2}{c|}{500} & \multicolumn{2}{c|}{500} & \\
            \midrule
            \multirow{2}{*}{\textbf{\makecell[c]{Cross-lingual}}} & \multicolumn{2}{c|}{\textbf{zh2en}} & \multicolumn{2}{c|}{\textbf{en2zh}} & \multirow{2}{*}{1,000} \\ \cmidrule(lr){2-5}
             & \multicolumn{2}{c|}{500} & \multicolumn{2}{c|}{500} & \\
            \bottomrule
            \end{tabular}
    \end{threeparttable}
}
\caption{Statistics of the proposed evaluation sets in four scenarios (\textbf{en}: English, \textbf{zh}: Chinese, \textbf{mixed}: mixture of English and Chinese, \textbf{zh2en}: Chinese reference speech with English target text. Similarly for \textbf{en2mixed}, \textbf{zh2mixed}, and \textbf{en2zh}).}
\label{tab:eval-data}
\end{center}
\end{table}

\subsection{Objective Evaluation Metrics}

For objective metrics, we evaluate the intelligibility (WER), speaker similarity (SIM), and overall speech quality (UTMOS~\cite{utmos}): 
\begin{itemize}[itemsep=0ex,leftmargin=2ex]
    \item \textbf{WER}: We employ \texttt{Whisper-large-v3}\footnote{\href{https://huggingface.co/openai/whisper-large-v3}{https://huggingface.co/openai/whisper-large-v3}}~\cite{whisper} for English texts, and \texttt{Paraformer-zh}\footnote{\href{https://huggingface.co/funasr/paraformer-zh}{https://huggingface.co/funasr/paraformer-zh}}~\cite{paraformer, funasr} for Chinese and code-switching texts.
    \item \textbf{SIM}: We compute the cosine similarity between the WavLM TDNN\footnote{\href{https://github.com/microsoft/UniSpeech/tree/main/downstreams/speaker_verification}{https://github.com/microsoft/UniSpeech/tree/main/\\downstreams/speaker\_verification}}~\cite{wavlm} speaker embeddings of generated samples and the prompt samples.
    \item \textbf{UTMOS}: We use the pretrained UTMOS strong learner following the official implementation\footnote{\href{https://github.com/sarulab-speech/UTMOS22}{https://github.com/sarulab-speech/UTMOS22}}.
\end{itemize}

\subsection{Subjective Evaluation}\label{sec:appendix-sub-eval}

We consider four different settings: regular, articulatory, code-switching, and cross-lingual. Each setting is evaluated in two languages, resulting in 10 samples per language. This setup yields a total of 80 pairs. These 80 pairs are evaluated across 5 different systems (ARS, F5-TTS, MaskGCT, CosyVoice 2, and Ints), leading to a total of 400 pairs. We engage 20 participants in the evaluation process, ensuring that each sample is assessed at least three times.

\begin{figure}[h]
    \centering
    \includegraphics[width=1\linewidth]{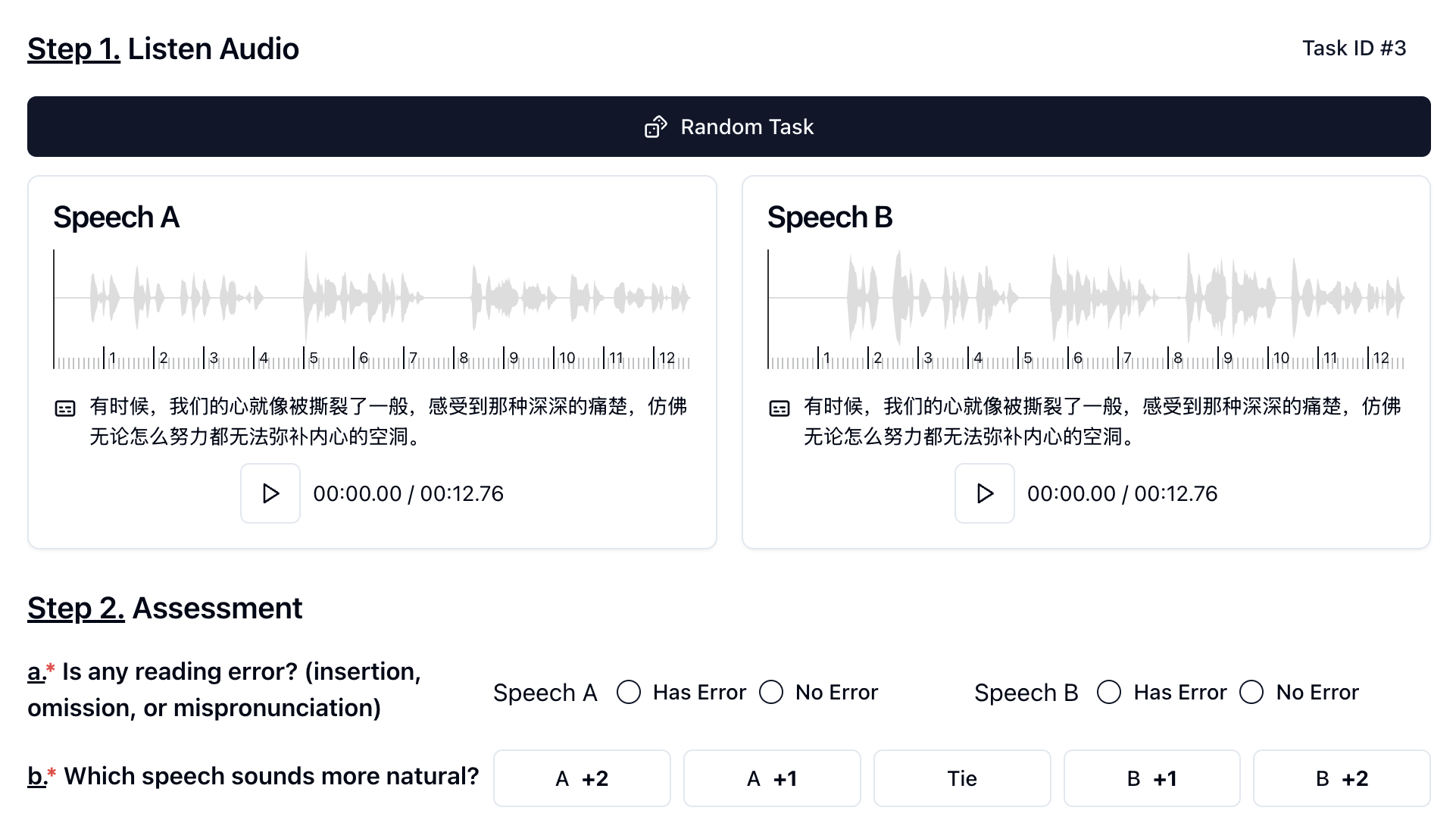}
    \caption{User interface for intelligibility and naturalness evaluation.}
    \label{fig:reading_err_and_natural}
\end{figure}

\begin{figure}[h]
    \centering
    \includegraphics[width=1\linewidth]{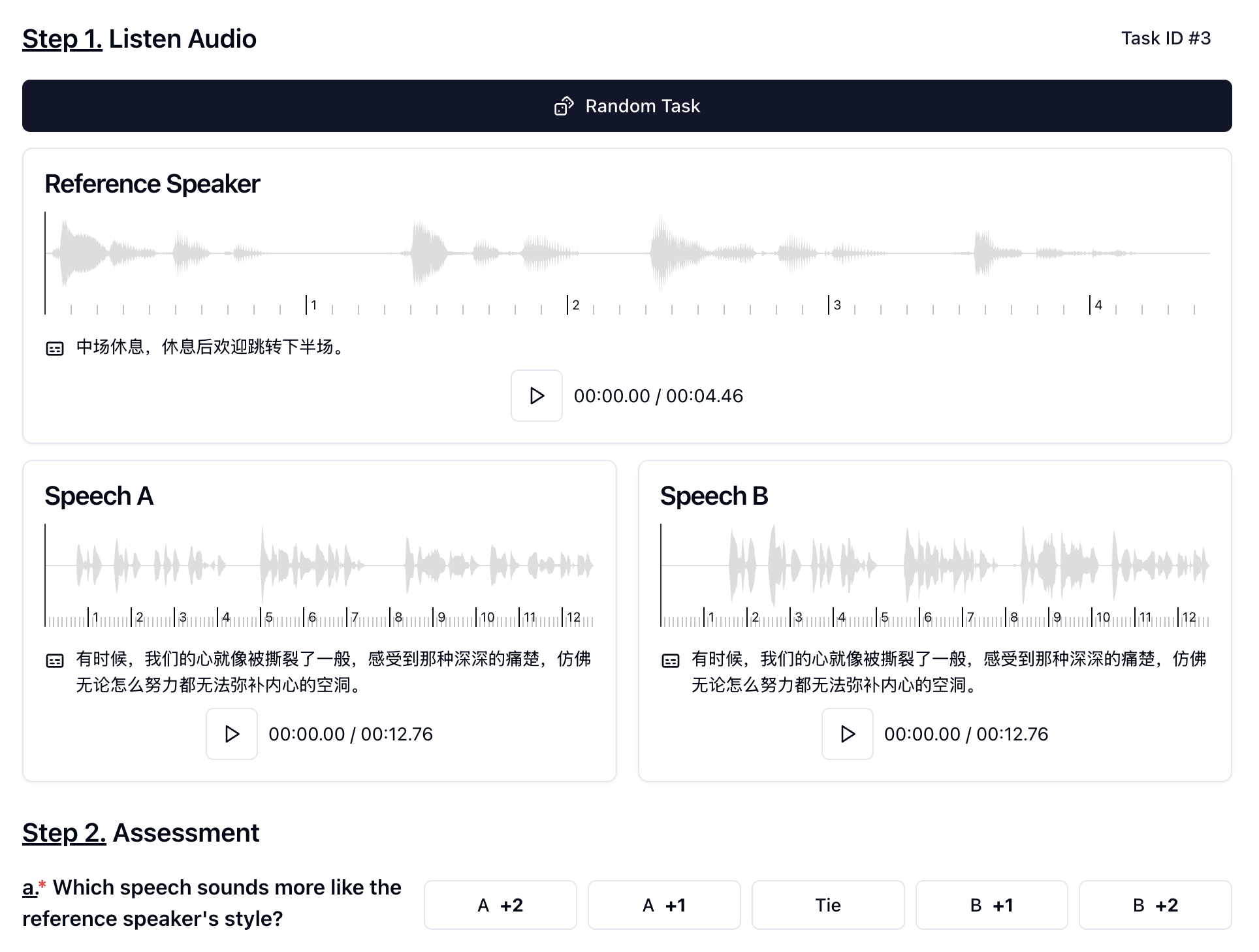}
    \caption{User interface for speaker similarity evaluation.}
    \label{fig:ref_audio}
\end{figure}

We conduct subjective evaluations from three perspectives: intelligibility (reading accuracy), naturalness (N-CMOS), and speaker similarity (A/B Testing). We have developed an automated subjective evaluation interface, as shown in Figure~\ref{fig:reading_err_and_natural} and Figure~\ref{fig:ref_audio}. For each item to be evaluated, users will see three components: the System Interface, the Questionnaire, and the Evaluation Criteria.

    \paragraph{Intelligibility (Reading Accuracy):}
    
    \begin{itemize}[itemsep=0ex,leftmargin=2ex]
        \item \textbf{System Interface:} Users listen to the speech audio and compare it to the provided target text to assess whether the speech matches the text.
        
        \item \textbf{Questionnaire:} Users are asked, ``Is any reading error? (insertion, omission, or mispronunciation)''

         \item \textbf{Evaluation Criteria:} The evaluation is binary: ``No Error'' (the speech matches the text) or ``Has Error'' (the speech does not match the text).
    \end{itemize}
    
    \paragraph{Naturalness (N-CMOS):}
    
    \begin{itemize}[itemsep=0ex,leftmargin=2ex]
        \item \textbf{System Interface:} Users listen to two speech samples, A and B, to compare their naturalness.
        
        \item \textbf{Questionnaire:} Users are asked, ``Which speech sounds more natural?''
        
        \item \textbf{Evaluation Criteria:} Options include A +2 (Sample A is much more natural), A +1 (Sample A is slightly more natural), Tie (Both are equally natural), B +1 (Sample B is slightly more natural), and B +2 (Sample B is much more natural).
    \end{itemize}
    
    \paragraph{Speaker Similarity (A/B Testing):}

    \begin{itemize}[itemsep=0ex,leftmargin=2ex]
        \item \textbf{System Interface:} Users listen to two speech samples, A and B, to evaluate their similarity to the speech of the reference speaker.
        
        \item \textbf{Questionnaire:} Users are asked, ``Which speech sounds more like the reference speaker's style?''
        
        \item \textbf{Evaluation Criteria:} Options include A +2 (Sample A is much more similar), A +1 (Sample A is slightly more similar), Tie (Both are equally similar), B +1 (Sample B is slightly more similar), and B +2 (Sample B is much more similar).
    \end{itemize}

\end{document}